\documentclass[fleqn,usenatbib]{mnras}
\usepackage[varg]{txfonts}
\usepackage[T1]{fontenc}
\usepackage{xcolor}

\DeclareRobustCommand{\VAN}[3]{#2}
\let\VANthebibliography\thebibliography
\def\thebibliography{\DeclareRobustCommand{\VAN}[3]{##3}\VANthebibliography}

\usepackage{graphicx}	
\usepackage{soul}

\newcommand{\T}[1]{\ensuremath{T_\mathrm{#1}}}
\newcommand{\Tb}  {\T{b}}
\newcommand{\Tbg} {\T{bg}}
\newcommand{\Tex} {\T{ex}}
\newcommand{\Th}  {\ensuremath{T\sub{k}\Sup{hom}}}
\newcommand{\Tk}  {\T{k}}
\newcommand{\Tl}  {\T{L}}

\newcommand{\Trot}{\T{rot}}
\newcommand{\kms} {km~s$^{-1}$}

\newcommand{\nh}  {NH$_3$}

\newcommand{\cmt} {cm$^{-3}$}

\newcommand{\phs}{\phantom{$-$}}

\newcommand{\equal} {\ensuremath{\!\!\!&=&\!\!\!}}      
\newcommand{\sub}[1]{\ensuremath{_\mathrm{#1}}}
\newcommand{\Sup}[1]{\ensuremath{^\mathrm{#1}}}
\newcommand{\mm}{\ensuremath{\mathrm{m}}}
\newcommand{\ms}{\ensuremath{\mathrm{s}}}
\newcommand{\hms}[4]{\ensuremath{#1^\mathrm{h}#2^\mathrm{m}#3\fs#4}}

\newcommand{\dmsf}[4]{\ensuremath{#1\degr#2'#3\farcs#4}}
\newcommand{\RA}[4] {$\alpha(J2000)=\hms{#1}{#2}{#3}{#4}$}

\newcommand{\DECf}[4]{$\delta(J2000)=\dmsf{#1}{#2}{#3}{#4}$}

\title[
NH$_3$ spherical model
]{
Can radial temperature profiles be inferred using NH$_3$ (1, 1) and (2, 2) observations?
}
\author[
R. Estalella, A. Palau, \& G. Busquet
]{
Robert Estalella,$^{1,2,3}$\thanks{E-mail: robert.estalella@gmail.com}
Aina Palau,$^{4}$
and
Gemma Busquet$^{1,2,3}$
\\
$^{1}$Departament de F\'{\i}sica Qu\`antica i Astrof\'{\i}sica (FQA), Universitat de Barcelona (UB), Mart\'{i} i Franqu\`es 1, 08028 Barcelona, Spain
\\
$^{2}$Institut de Ci\`encies del Cosmos (ICCUB), Universitat de Barcelona (UB), Mart\'{i} i Franqu\`es 1, 08028 Barcelona, Spain
\\
$^{3}$Institut d'Estudis Espacials de Catalunya (IEEC), Gran Capit\`a, 2-4, 08034 Barcelona, Spain
\\
$^{4}$Instituto de Radioastronom\'{\i}a y Astrof\'{\i}sica, Universidad Nacional Aut\'onoma de M\'exico, Antigua Carretera a P\'atzcuaro 8701, Ex-Hda.\ San Jos\'e de la Huerta,  \\
58089 Morelia, Michoac\'an, M\'exico 
}

\date{Accepted 2024 January 18. Received 2024 January 11 ; in original form 2023 March 08}
\pubyear{2023}
\begin{document}
\label{firstpage}
\pagerange{\pageref{firstpage}--\pageref{lastpage}}
\maketitle

\begin{abstract}
A number of works infer radial temperature profiles of envelopes surrounding young stellar objects using several rotational transitions in a pixel-by-pixel or azimuthally-averaged basis. 
However, in many cases the assumption that the rotational temperature is constant along the line of sight is made, while this is not the case when a partially resolved envelope, assumed to be spherically symmetric, is used to obtain values of temperature for different projected radii. 
This kind of analysis (homogeneous analysis) is intrinsically inconsistent. 
By using a spherical envelope model to interpret \nh{} $(1,1)$ and $(2,2)$ observations, we tested how robust it is to infer radial temperature profiles of an envelope. 
The temperature and density of the model envelope are power laws of radius, but the density can be flat for an inner central part. 
The homogeneous analysis was applied to obtain radial temperature profiles, and resulted that for small projected radii, where the optical depth of the lines is high, the homogeneous temperature can be much higher than the actual envelope temperature. 
In general, for larger projected radii, both the temperature and the temperature power-law index can be underestimated by as much as 40\%, and  0.15, respectively.
We applied this study to the infrared dark cloud G14.225$-$0.506 for which the radial temperature profile was previously derived from the dust emission at submillimeter wavelengths and the spectral energy distribution.
As expected, the homogeneous analysis underestimated both the temperature and the temperature power-law index.
\end{abstract}


\section{Introduction}

Characterizing the envelopes surrounding young stellar objects is important to correctly infer their properties and compare them to theoretical models. 
A typical assumption used to describe these envelopes is to consider that their density and temperature decrease with radius following power-laws. This has widely been determined from observations \citep[e.g.,][]{Beu02,Mue02,Hat03,Will05,But12,Gia13,Pal14,Pal21,Gie21} and is also predicted from theoretical models \citep[e.g.,][]{Lar69,Shu77,Gom21}. 

The density and temperature structure of the envelopes surrounding young stellar objects have been determined in many cases using the radial intensity profiles from the continuum emission. 
Even though it is widely known that molecular clouds are largely filamentary, it is also true that the most active star-forming sites are located at the centers of converging filaments, giving rise to the so-called hub-filament systems \citep[e.g.,][]{Dewangan20,Wang20,Kumar22,Liu23}. It has been shown that the hubs are reasonably reproduced by models of spherical symmetry, such as those studied in  \citet{Beuther23}. Actually, \cite{Kainulainen13} find that fragmentation of molecular clouds at spatial scales 0.5--10 pc is in agreement with the fragmentation expected for a self-gravitating cylinder, and therefore a filamentary structure, while at scales $<0.5$~pc the fragmentation is in agreement with spherical Jeans fragmentation. 
In addition, it has been shown that assuming spherical geometry is reasonable for cores embedded within filamentary structures for aspect ratios up to 2 \citep{Roy14}.
Thus, even in the era of filamentary molecular clouds, assuming spherical symmetry to model the hubs where active star formation is taking place seems a reasonable first approximation.

In the case of spherical symmetry, and if one assumes optically thin emission, the Rayleigh-Jeans approximation, and an envelope with infinite radius, the intensity as function of the projected distance to the source, $b$, follows a power-law with index $1-(p+q)$, where $p$ and $q$ are the power-law indices of the density and temperature \citep[e.g.,][]{Bel02,Beu02,Pal14}. 
Therefore, given an observed radial intensity profile, one needs to make a specific assumption about either the density or the temperature to infer the other.

Molecular line observations, including transitions of \nh, H$_2$CO, CH$_3$CN, HCOOCH$_3$, and complex organic molecules, of partially resolved envelopes have been used to derive the temperature radial profile of the envelope \citep[e.g.,][]{Ahm18,Bel18,Gie19,Gie21,Busch22}.
The standard procedure to infer the temperature from rotational diagrams, for instance, of \nh{}, assumes that, along the line of sight, the medium is homogeneous and in particular that the rotational temperature is constant.
However, the derived values of $\Trot$ for different projected radii are interpreted as the radial profile of temperature of the envelope. 
This kind of analysis is inconsistent because the gas is assumed to be homogeneous along the line of sight and, at the same time, inhomogeneous in the radial direction.

In this paper, we aim at studying whether the determination of the radial temperature profiles from the \nh{} $(1,1)$ and $(2,2)$ data are a reasonable approach to the actual temperature profile. 
Since the collisional coefficients for the quadrupole-hyperfine transitions of \nh{} are not available in the literature, the existing numerical radiative transfer models are of limited usefulness. 
Thus, we developed a numerical radiative transfer calculation for the \nh{} $(1,1)$ and $(2,2)$ hyperfines, which is described in Appendix A. A comparison with a full radiative transfer calculation is also included.
The structure of the paper is as follows:
in Section 2 we present the basics of the standard \nh{} analysis, 
in Section 3 we present the limitations of this analysis, in Section 4 the spherical model used here is described, 
in Section 5 the optically thin case is considered, 
in Section 6 a radiative transfer calculation is used for different parameters of the envelope, 
in Section 7 the results found here are applied to the G14.225$-$0.506 case and 
in Section 8 we summarize our main conclusions.

\section{\nh{} $(J,K)= (1,1)$ and $(2,2)$ analysis basics}
\label{sec:basics}

\begin{figure}
\centering
\includegraphics[width=0.8\columnwidth]{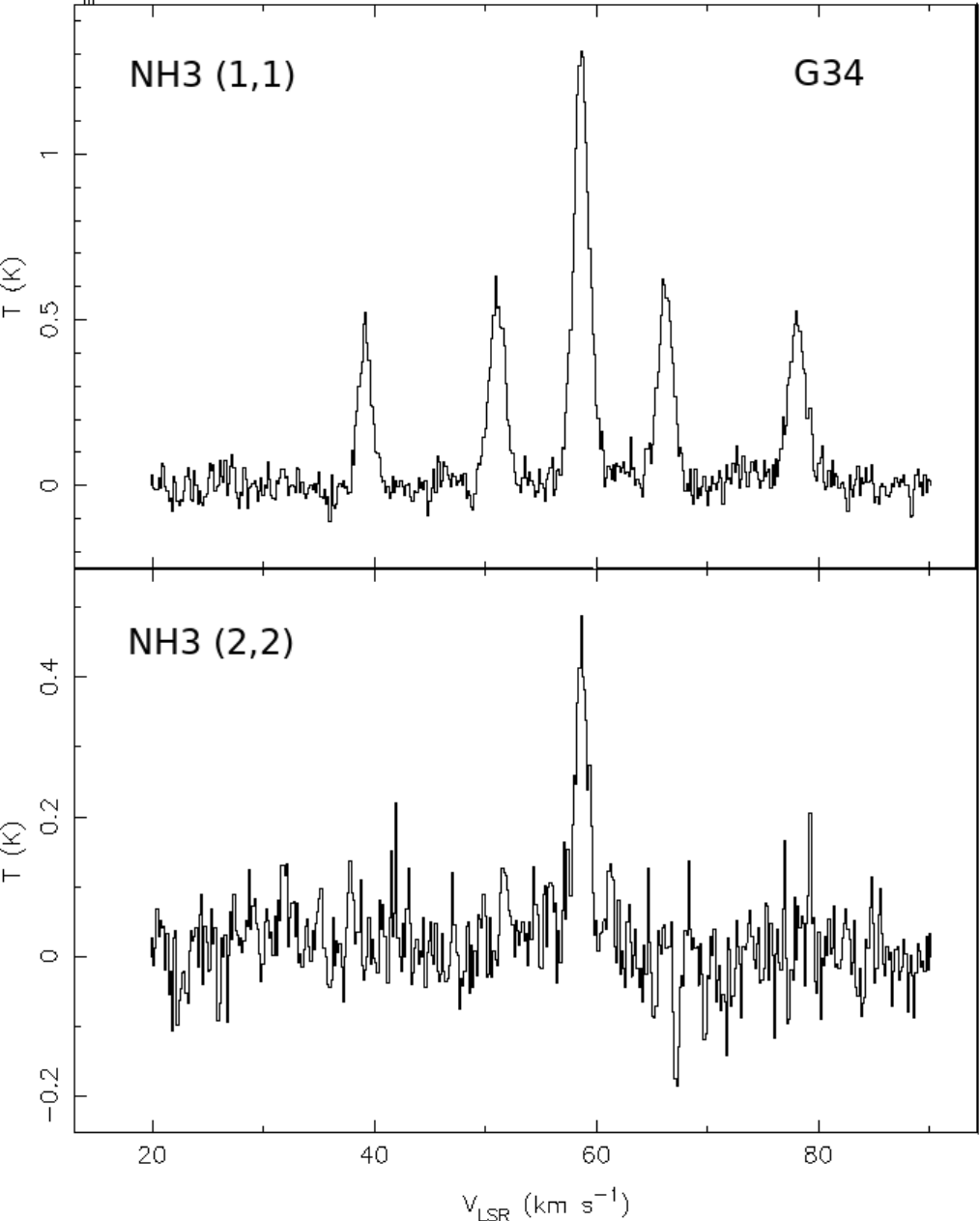}
\caption{\label{fig:g34}
Example of \nh{} $(J,K)= (1,1)$ (top) and $(2,2)$ (bottom) lines in G34.43+0.24. The $(1,1)$ transition shows the main and four (two inner and two outer) satellite quadrupole hyperfine lines, while the satellite lines of the $(2,2)$ transition are too faint to be observed.
}
\end{figure}

The intensity of the \nh{} $(J,K)= (1,1)$ and $(2,2)$ lines can be expressed as
\begin{equation}\label{eq:Tjk}
\Tl(j,k) =  \eta\sub{B} [J_\nu(\Tex)-J_\nu(\Tbg)] (1-\exp[-\tau(j,k)]),
\end{equation}
where 
$\eta\sub{B}$ is the beam filling-factor,
$J_\nu(T)$ is the Plank correction to the Rayleigh-Jeans approximation,
\begin{equation}
J_\nu(T)= \frac{h\nu_{jk}/k}{\exp(h\nu_{jk}/kT)-1}, 
\end{equation}
$\nu_{jk}$ is the frequency of the transition,
$\Tex$ is the excitation temperature, 
$\Tbg$ is the background temperature,
and
$\tau(j,k)$ is the optical depth of the transition.
Since we are dealing with the observation of an angularly resolved envelope, in the following we will consider the beam filling factor $\eta\sub{B}= 1$.

The standard procedure to analyze observations of the $(1,1)$ and $(2,2)$ inversion transitions of ammonia consists in fitting the intensity of the main and inner-satellite quadrupole-hyperfines of the $(1,1)$ line (see Fig \ref{fig:g34}), $\Tl(1,1;\mm)$ and $\Tl(1,1;\ms)$. Since the optical depth ratio of the main and inner satellite lines is known $R_{jk\mm}/R_{jk\ms}=18/5=3.6$ (see Table \ref{tab:nh3m}), the optical depth of the main line, $\tau(1,1;\mm)$ is obtained from the equation
\begin{equation}\label{eq:T1ratio}
\frac{\Tl(1,1;\mm)}{\Tl(1,1;\ms)}= \frac{1-\exp[-\tau(1,1;\mm)]}{1-\exp[-\tau(1,1;\mm)/3.6]}.
\end{equation}
This approach is valid if the magnetic hyperfines are blended, that is, the hyperfines line-width is large enough so that the quadrupole hyperfines profiles show no substructure and are nearly Gaussian. 

A more robust method is to fit the magnetic hyperfine structure of the $(1,1)$ line using a procedure like CLASS (NH$_3$ method). 
The outputs of the fit include the intensity and optical depth of the main line, $\Tl$$(1,1;\mm)$ and $\tau$$(1,1;\mm)$.
For the $(2,2)$ line, the satellite lines are usually unobservable (see Fig.\  \ref{fig:g34}), and a single Gaussian is fitted to the main line, giving the main line intensity, $\Tl(2,2;\mm)$.
Since the excitation temperature of the $(1,1)$ and $(2,2)$ transitions are assumed to be the same, and the frequencies of the two transitions are very close, the intensity ratio can be used to derive the optical depth of the $(2,2)$ main line, from the relation
\begin{equation}\label{eq:tau22}
1-\exp[-\tau(2,2;\mm)]= \frac{\Tl(2,2;\mm)}{\Tl(1,1;\mm)} \, \left(1-\exp[-\tau(1,1;\mm)]\right).
\end{equation}

Alternatively, the magnetic hyperfine structure of the $(1,1)$ and $(2,2)$ lines can be fitted simultaneously using a procedure like HfS \citep{Est17}, and the optical depth of the $(2,2)$ main line is obtained directly, with the same assumption of equal excitation temperature for the $(1,1)$ and $(2,2)$ lines
\citep[see for instance][]{Sep20}.

\begin{table}
\centering
\caption{
Frequencies in temperature units,
spontaneous emission coefficients, 
and 
optical depth ratio of the main (m), inner satellite (s) of \nh{} $(1,1)$ and $(2,2)$ to the total optical depth of the transition  \citep{Oso09, Man15}.
\label{tab:nh3m}}
\begin{tabular}{lcccc}
\hline
     & $h\nu_{jk}/k$ & $A_{jk}$    & $R_{jkm}=$              & $R_{jks}=$ \\
Line & (K)  & (10$^{-7}$ s$^{-1}$) & $\tau(j,k;\mm)/\tau(j,k)$ & $\tau(j,k;\ms)/\tau(j,k)$\\
\hline
$(1, 1)$  & 1.13716 & 1.66838 & $1/2$            & $5/36=0.1389$ \\
$(2, 2)$  & 1.13850 & 2.23246 & $43/54=0.7963$   & $7/135=0.0519$ \\
\hline
\end{tabular}
\end{table}

Once the opacities are known, Eq.\  \ref{eq:Tjk} is used to derive the excitation temperature $\Tex$. 
This temperature is not well determined since it depends on the correct calibration of telescope output, and on the coupling of the telescope beam with the source (the beam filling factor). 
The usual assumption is to take $\eta\sub{B}=1$. 
However, in the Rayleigh-Jeans limit and the optically thin case, the \nh{} column density can be derived independently of the value of $\Tex$ (see Section \ref{sec:thin}).

From these data, the column density of ammonia molecules in the $(1,1)$ and $(2,2)$ levels, $N_{11}$ and $N_{22}$, can be derived \citep[see for instance,][]{Est17},
\begin{equation}\label{eq:Njk}
N_{jk}= 
\sqrt{\frac{\pi}{4\ln2}} \,
\frac{8\pi\nu_{jk}^3}{c^3A_{jk}}\, 
\frac{\exp(h\nu_{jk}/k\Tex)+1}{\exp(h\nu_{jk}/k\Tex)-1}\,
\frac{\tau(j,k;\mm)}{R_{jk\mm}} \, \Delta v,
\end{equation}
where
$A_{jk}$ is the spontaneous emission coefficient (Table \ref{tab:nh3m}), 
$R_{jk\mm}= \tau(j,k;\mm)/\tau_{jk}$, is the ratio of main line to total optical depth, given in Table \ref{tab:nh3m}, and
$\Delta v$ is the full-width at half-maximum of the hyperfine lines.

\begin{table}
\centering
\caption{
Degeneracies and
energies above the $(1,1)$ level
of the lower metastables levels of \nh{} \citep{Man15}.
\label{tab:nh3}}
\begin{tabular}{ccc}
\hline
        &          & $(E_{jk}-E_{11})/k$ \\
$(J,K)$ & $g_{jk}$ & (K) \\
\hline
$(0,0)$ & \phantom{1}4 & $          -22.64$ \\
$(1,1)$ & 12           & $\phantom{-2}0.00$ \\
$(2,2)$ & 20           & $\phantom{-}40.99$ \\
$(3,3)$ & 56           & $\phantom{-}99.76$ \\ 
\hline
\end{tabular}
\end{table}

To obtain the rotational temperature, $\Trot$, and the total \nh{} column density, $N(\mathrm{NH_3})$, it is necessary to determine the fraction, $f_{jk}$, of ammonia molecules that are in the $(J,K)$ rotational level,
\begin{equation}\label{eq:fjk}
n_{jk} = f_{jk}\,n(\mathrm{NH_3}),
\end{equation}
from which the total hydrogen gas density is derived adopting an ammonia fractional abundance, $X$, assumed to be constant,
\begin{equation}\label{eq:X}
n(\mathrm{NH_3}) = X\,n(\mathrm{H_2}).
\end{equation}

The usual assumption for moderate temperatures is that only the metastable levels $(J=K)$ of ammonia are populated, and that they are populated according to a unique excitation temperature, called the rotational temperature $\Trot$. 
The rotational temperature is given by the Boltzmann equation,
\begin{equation}\label{eq:fracff}
\frac{n_{22}}{n_{11}}= \frac{g_{22}}{g_{11}} \exp[-(E_{22}-E_{11})/k\Trot], 
\end{equation}
where $g_{jk}$ is the degeneracy and $E_{jk}$ the energy of the level $(J,K)$ (see Table \ref{tab:nh3}).
The latter equation is used routinely to derive the rotational temperature by assuming that $N_{22}/N_{11} \simeq n_{22}/n_{11}$,
\begin{equation}\label{eq:trot}
\Trot= \frac{(E_{22}-E_{11})/k}
{\displaystyle\ln\left(\frac{g_{22}}{g_{11}}\frac{N_{11}}{N_{22}}\right)}.
\end{equation}
The rotational temperature can also been expressed as a function of the optical depths of the $(1,1)$ and $(2,2)$ main lines by using Eq.\  \ref{eq:Njk},
\begin{equation}
\Trot= \frac{(E_{22}-E_{11})/k}
{\displaystyle\ln\left(
\frac{A_{22}\,R_{22\mm}\,g_{22}}{A_{11}\,R_{11\mm}\,g_{11}}
\frac{\tau(1,1;\mm)}{\tau(2,2;\mm)}
\right)},
\end{equation}
resulting, in practical units, in
\begin{equation}\label{eq:trot_tau}
\left[\frac{\Trot}{\mathrm{K}}\right]=
\frac{40.99}{\displaystyle\ln\left(3.544
\frac{\tau(1,1;\mm)}{\tau(2,2;\mm)}\right)}.
\end{equation}

The rotational temperature is a good approximation to the kinetic temperature if the radiative transitions up the $K$-ladder are negligible, and only collisional transitions between the metastable levels occur. 
The rotational temperature can be corrected to estimate the kinetic temperature \citep{Taf04}. 
An improved version of the correction is given in \citet{Est17}, 
\begin{equation}\label{eq:tk_trot}
\Trot= \frac{\Tk}{1+\displaystyle\frac{\Tk}{40.99} \ln \left[1+0.73 \exp (-16.26/\Tk)\right]}.
\end{equation}
The last equation has to be solved iteratively to obtain \Tk.

\begin{figure}
\includegraphics[width=\columnwidth]{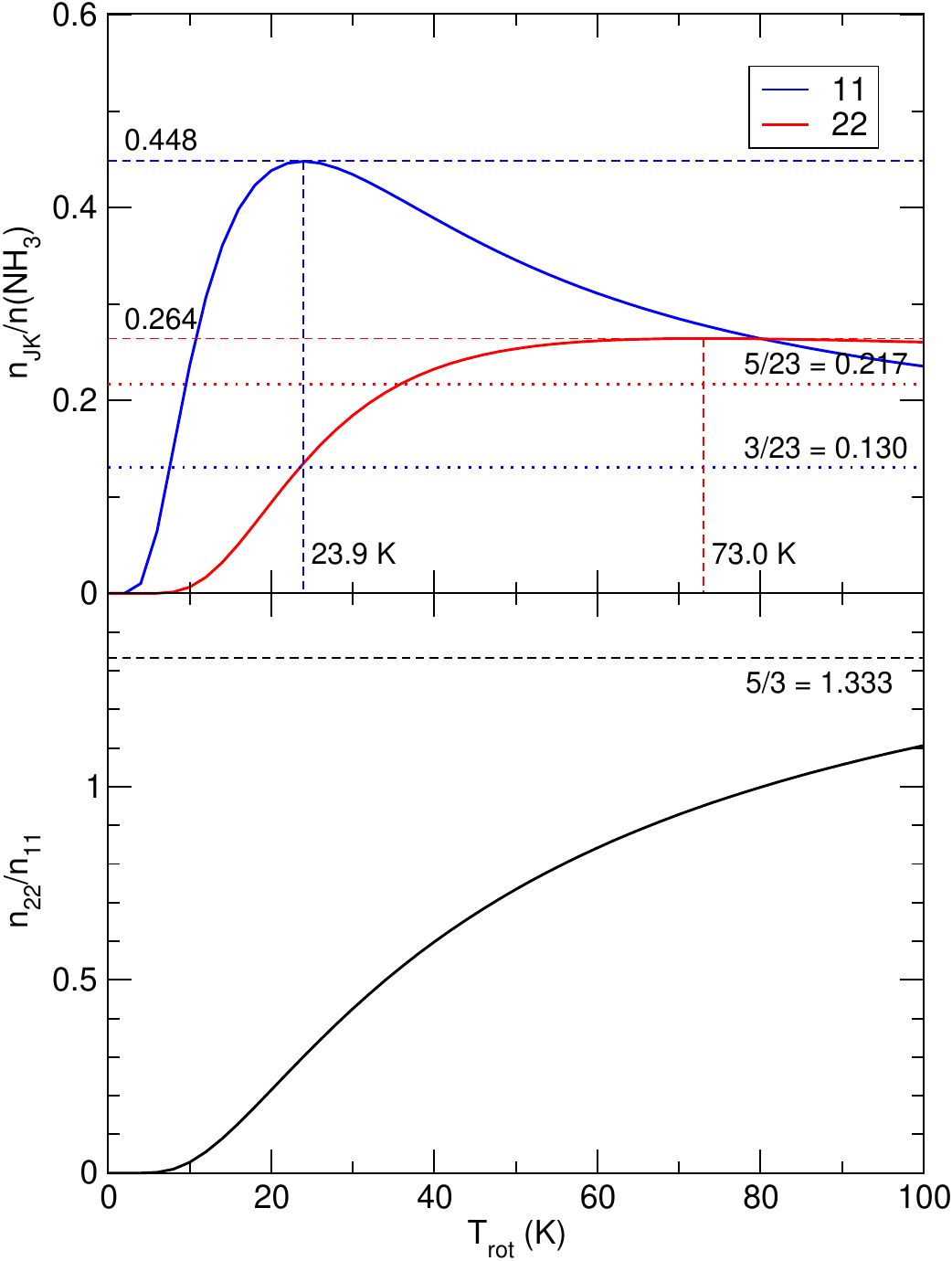}
\caption{\label{fig:njk_t}
Fractional abundance of NH$_3$ $(1,1)$ (blue) and $(2,2)$ (red) \emph{(top)} and ratio $n_{22}/n_{11}$ \emph{(bottom)} as a function of rotational temperature.}
\end{figure}

In order to obtain the total \nh{} column density, we can consider that 
for moderate values of $\Trot$ ($\la100$ K), only the levels $(J,K)= (0,0)$, $(1,1)$, $(2,2)$, and $(3,3)$ are populated, so that the partition function $Q(\Trot)$ is 
\begin{equation}
Q(\Trot)=   \sum_{jk=0,0}^{3,3} g_{jk}e^{-(E_{jk})/k\Trot}.
\end{equation}
The fractional abundances of the $(1,1)$ and $(2,2)$ levels, $f_{jk}=n_{jk}/n(\mathrm{NH_3})$, are given by
\begin{eqnarray}\label{eq:f11f22}
f_{11} \equal
\frac{g_{11}}{Q} e^{-(E_{11}/k\Trot)}= 
\frac{g_{11}}{\displaystyle\sum_{jk=0,0}^{3,3} g_{jk}e^{-(E_{jk}-E_{11})/k\Trot}}, \\
\label{eq:f22}
f_{22} \equal  
\frac{g_{22}}{Q} e^{-(E_{22}/k\Trot)}= 
\frac{g_{22}}{g_{11}} e^{-(E_{22}-E_{11})/k\Trot} f_{11}. 
\end{eqnarray}
The maximum of $f_{11}$ ($f_{11}\simeq 0.448$) occurs for $\Trot\simeq23.9$ K, and for high temperatures $f_{11}\to3/23=0.130$. 
The maximum  of $f_{22}$ ($f_{22}\simeq 0.264$) occurs for $\Trot\simeq73.0$ K, and for high temperatures $f_{22}\to5/23=0.217$ (Fig.\ \ref{fig:njk_t}, top). 
The ratio $f_{22}/f_{11}=n_{22}/n_{11}$ is a simple function of $\Trot$ (Eq. \ref{eq:f22}) and is shown in Fig.\ \ref{fig:njk_t} (bottom). 
It is an increasing function of $\Trot$, and for high temperatures $f_{22}/f_{11}\to5/3=1.333$.

Once $\Trot$ is known, the fractional abundance $f_{11}$ can be calculated, and we further assume that 
\begin{equation}
\frac{N_{11}}{N(\mathrm{NH_3})} \simeq \frac{n_{11}}{n(\mathrm{NH_3})}= f_{11},
\end{equation}
so the the hydrogen column density is 
\begin{equation}
N(\mathrm{H_2})= \frac{N_{11}}{X\,f_{11}}.
\end{equation}

\section{Limitations of the standard \nh{} analysis}
\label{sec:limitations}

There is an implicit assumption in the approximations made to derive the expressions derived so far: that the gas is homogeneous and, in particular, that the rotational temperature is constant along the line of sight. 
This has no importance as long as the clump of gas analyzed is assumed to be  homogeneous. However, this is not the case when we analyze ammonia data of a partially resolved envelope, assumed to be spherically symmetric, and  use Eq.\ \ref{eq:trot} to obtain values of $\Trot$ for different projected radii (the radial profile of temperature of the envelope). 
This kind of analysis is intrinsically  inconsistent. On the one hand, Eq.\ \ref{eq:trot} assumes that the gas is homogeneous along the line of sight, and on the other hand, these results are used to infer a radial dependence of $\Trot$, implying that $\Trot$ is not constant along the line of sight.

In addition, the assumption of homogeneity along the line of sight can be incompatible with the observed intensities of the lines when observing a region with strong gradients of density or temperature along the line of sight. 
Let us examine the three equations used to derive 
$\tau(1,1;\mm)$ (Eq.\  \ref{eq:T1ratio}), 
$\tau(2,2;\mm)$, (Eq.\  \ref{eq:tau22}), and 
$\Trot$ (Eq.\  \ref{eq:trot_tau}).
For simplicity, in this discussion we will call
$\tau_{1\mm}= \tau(1,1;\mm)$,
$\tau_{2\mm}= \tau(2,2;\mm)$, 
$T_{1\mm}   =  \Tl(1,1;\mm)$,
$T_{1\ms}   =  \Tl(1,1;\ms)$, 
$T_{2\mm}   =  \Tl(2,2;\mm)$, and
$r_{21}$ the ratio of $(2,2)$ to $(1,1)$ intensities, $r_{21}=T_{2\mm}/T_{1\mm}$.

Let us call $r_{11}$ the intensity ratio of main to satellite $(1,1)$ lines, $r_{11}=T_{1\mm}/T_{1\ms}$.
Equation \ref{eq:T1ratio} used to derive the optical depth of the $(1,1)$ line requires that 
\begin{equation}\label{eq:cond_tau1}
1 < r_{11} < \frac{R_{11\mm}}{R_{11\ms}} = 3.6.
\end{equation}
For $r_{11}\to1$ we have $\tau_{1\mm}\to+\infty$,
while for $r_{11}=3.6$, the optical depth is zero.
If $r_{11}$ is less than 1, no value of the optical depth $\tau_{1\mm}$ can be derived, and if it is greater than 3.6, Eq.\ \ref{eq:T1ratio} gives a negative optical depth. 

The derivation of the rotational temperature (Eq.\ \ref{eq:trot_tau}) requires that 
\begin{equation}\label{eq:cond_Trot}\label{eq:atau1}
\tau_{2\mm} < a \,\tau_{1\mm},
\end{equation}
where $a= (A_{22}\,R_{22\mm}\,g_{22})/(A_{11}\,R_{11\mm}\,g_{11}) = 3.544$.
The relation between $T_{1\mm}$, $T_{2\mm}$,  $\tau_{1\mm}$, and $\tau_{2\mm}$ (Eq.\ \ref{eq:tau22}) can be written as
\begin{equation}
\frac{1-\exp(-\tau_{1\mm})}{1-\exp(-\tau_{2\mm})} \, r_{21} = 1.
\end{equation}
By substituting  Eq.\ \ref{eq:atau1}, we obtain a condition on the intensity ratio $r_{21}$ and $\tau_{1\mm}$,
\begin{equation}\label{eq:cond_rtau}
r_\tau= \frac{1-\exp(-\tau_{1\mm})}{1-\exp(-a\,\tau_{1\mm})} \, r_{21} < 1.
\end{equation}
The values of $r_\tau$ for the limiting values of $\tau_{1\mm}$ are 
$r_\tau\to r_{21}/a$ for $\tau_{1\mm}\to0$, and 
$r_\tau\to r_{21}$ for $\tau_{1\mm}\to\infty$.

\begin{figure}
\includegraphics[width=0.9\columnwidth]{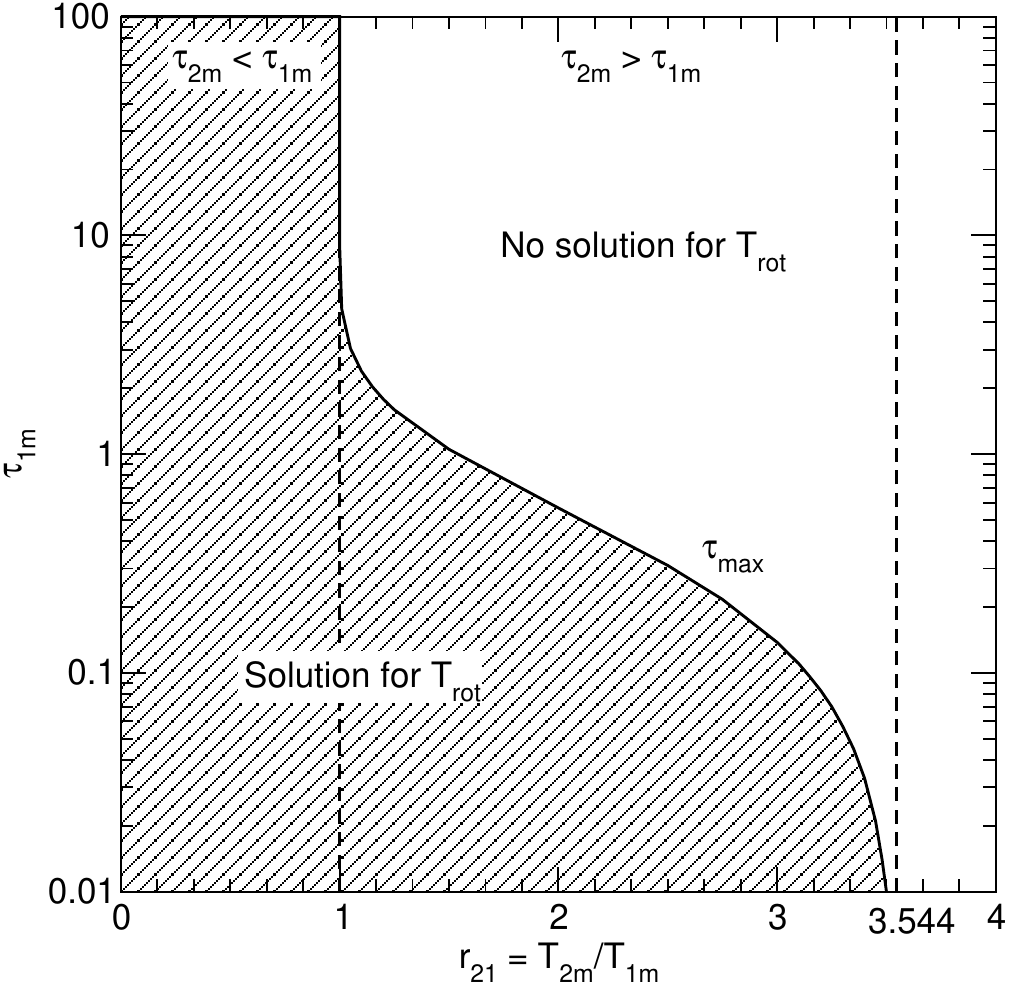}
\caption{\label{fig:taumax}
Plot of 
$\tau\sub{max}$ (Eq.\ \ref{eq:taumax}), 
as a function of the intensity ratio $r_{21}=\Tl(2,2;\mm)/\Tl(1,1;m)$, 
for $1 < r_{21} < 3.544$ (see text).
A value of $\Trot$ can only be obtained for $r_{21}<1$, or for $r_{21}>1$ and $\tau(1,1;\mm)<\tau\sub{max}$ (hatched area).
}
\end{figure}

In order to fulfill Eq.\ \ref{eq:atau1}, the intensity ratio $r_{21}$ and $(1,1)$ optical depth $\tau_{1\mm}$ can not be arbitrary. 
The intensity ratio must be $r_{21}<a=3.544$. 
If $r_{21}<1$ (that is, $(2,2)$ intensity lower than $(1,1)$), the optical depth can have any value. However, if $r_{21}>1$, $\tau_{1\mm}$ must be below a maximum value $\tau\sub{max}$, solution of the equation, for $r_{21}>1$,
\begin{equation}\label{eq:taumax}
\frac{1-\exp(-\tau\sub{max})}{1-\exp(-a\tau\sub{max})} \, r_{21} = 1.
\end{equation}
For $r_{21}\to1$, the solution of the equation gives $\tau\sub{max}\to\infty$, while for $r_{21}\to a$, we have $\tau\sub{max}\to0$ (see Fig.\ \ref{fig:taumax}).

In summary, in order to obtain a value of the rotational temperature, the  observational parameters,  
$r_{11}= T_{1\mm}/T_{1\ms}$, and 
$r_\tau=r_{21} [1-\exp(-\tau_{1\mm})]/[1-\exp(-a\,\tau_{1\mm})]$,
where 
$r_{21}=T_{2\mm}/T_{1\mm}$ and
$a= 3.544$,
must fulfill the two conditions: 
\begin{description}
\item[{$(1,1)$ line only:} ] $1\le r_{11} \le 3.6$ , and
\item[{$(1,1)$ and $(2,2)$ lines:} ] $r_\tau\le 1$.
This condition is fulfilled either if $r_{21}<1$ or, if $1<r_{21}<a$, for $\tau_{1\mm}<\tau\sub{max}$,
where $\tau\sub{max}$ is the solution of Eq.\ \ref{eq:taumax}.
\end{description}

In the following we will call the standard \nh{} analysis, under the assumption of gas homogeneity, \emph{homogeneous analysis}, and the kinetic temperature derived from Eqs.\ \ref{eq:trot} and \ref{eq:tk_trot}, \emph{homogeneous temperature}, $\Th$. 
We will estimate $\Th$, in a consistent way, for a spherically symmetric envelope, and its dependence on the density and temperature structure of the envelope.

\section{Spherical envelope model}

\begin{figure}
\centering
\includegraphics[width=0.9\columnwidth]{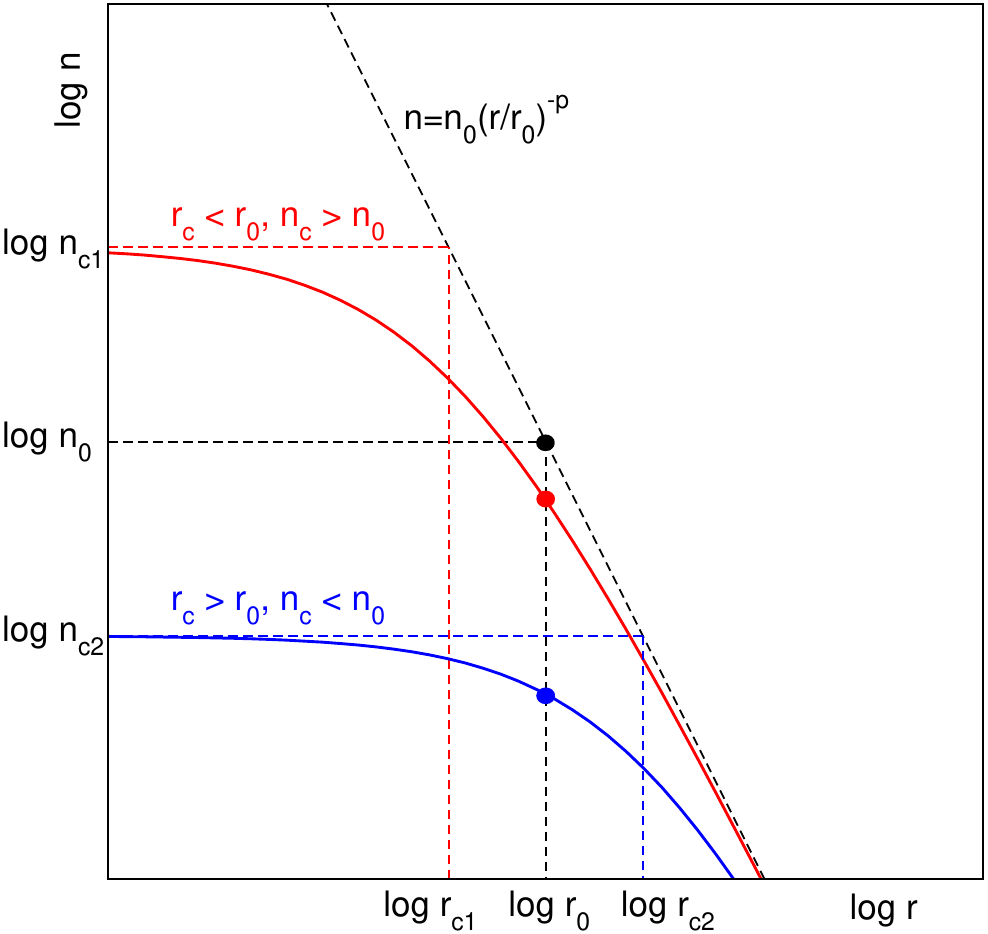}
\caption{Plummer-like density as a function of radius for the same values of $r_0$, $n_0$, and $p$, and two different values of $r_c$, one lower (red lines) and the other higher (blue lines) than $r_0$.
The red and blue dots indicate the actual density of the envelope at $r_0$, $n(r_0)$, while the black dot indicates the density parameter $n_0$ (see text).
}
\label{fig:plummer}
\end{figure}

Let us assume a spherically symmetric envelope, with a density that at large radii is a power-law of radius with index $-p$,
\begin{equation}\label{eq:n_power-law}
n(r) = n_0 \, (r/r_0)^{-p},
\end{equation}
but that may have a flat inner part, the so-called Plummer-like density, with a characteristic radius $r_c$,
\begin{equation}\label{eq:plummer}
n(r) = n_c\,\left[1+(r/r_c)^2\right]^{-p/2}.
\end{equation}
The central density, at $r=0$, is $n_c$.
For small radii, $r\ll r_c$, the density is flat, $n\simeq n_c$, while for large radii, $r\gg r_c$, the density follows a power law, $n\simeq n_c (r/r_c)^{-p}$.
We can consider the Plummer-like density as a generalization of the power-law density. 
This can be seen by writing the Plummer-like function as
\begin{equation}\label{eq:plummer2}
n(r) = n_0 \left[(r_c/r_0)^2+(r/r_0)^2\right]^{-p/2}.
\end{equation}
This expression of the Plummer-like function is useful because for $r_c=0$ we obtain directly a power-law function, with the reference radius $r_0$ and the density parameter $n_0$ well defined, while on the contrary, the central density $n_c\to\infty$.
In the small and large radii limits we obtain
\begin{eqnarray}
(r\ll r_c) & & n(r) \simeq n_0 \, (r_c/r_0)^{-p},
\nonumber \\
(r\gg r_c) & & n(r) \simeq n_0 \, (r/r_0)^{-p}. 
\end{eqnarray}
Thus, the central density $n_c$ and $n_0$ are related through
\begin{equation}
n_c= n_0 \, (r_c/r_0)^{-p}.  
\end{equation}
The density parameter $n_0$ is not an actual density. 
It can be interpreted as the density at the reference radius $r_0$ of the extrapolation of the Plummer-like function for $r\gg r_c$ (see Fig.\ \ref{fig:plummer}).
In the case $r_c>r_0$, $n_0$ is even higher than the maximum density of the envelope, $n_c$.

We consider that the Plummer-like law for the density is valid up to the envelope radius, taken as the radius for which the density is a given minimum value, $n\sub{min}$, the ambient gas density.

Regarding the temperature of the envelope, we consider a a power law of the radius, with a power-law index $-q$,
\begin{equation}\label{eq:T_power-law}
T(r) = T_0 (r/r_0)^{-q},\\
\end{equation}
where $T_0$ is the temperature at the reference radius $r_0$.
The gas temperature has also a minimum value, $\T{min}$, which corresponds to a radius such that for larger radii the envelope temperature is taken constant, equal to $\T{min}$.
The temperature power-law index for a centrally heated envelope, with the dust and the gas well coupled, depends on the dust opacity index $\beta$, $q=2/(4+\beta)$. 
For the usual values of $\beta$, $0<\beta<4$, the temperature power-law index results in $0.25<q<0.5$.

In the following we will derive the homogeneous temperature, $\Th$, for a spherical envelope as a function of the projected radius, $b$.
Two different approximations will be considered:
\begin{itemize}
\item[(i) ] Rough approximation, optically thin case, power-law density.
\item[(ii)] Radiative transfer calculation, with an excitation temperature given by the two-levels model, Plummer-like density. 
\end{itemize}

\section{Optically thin case, power-law density}
\label{sec:thin}

\begin{figure}
\includegraphics[width=\columnwidth]{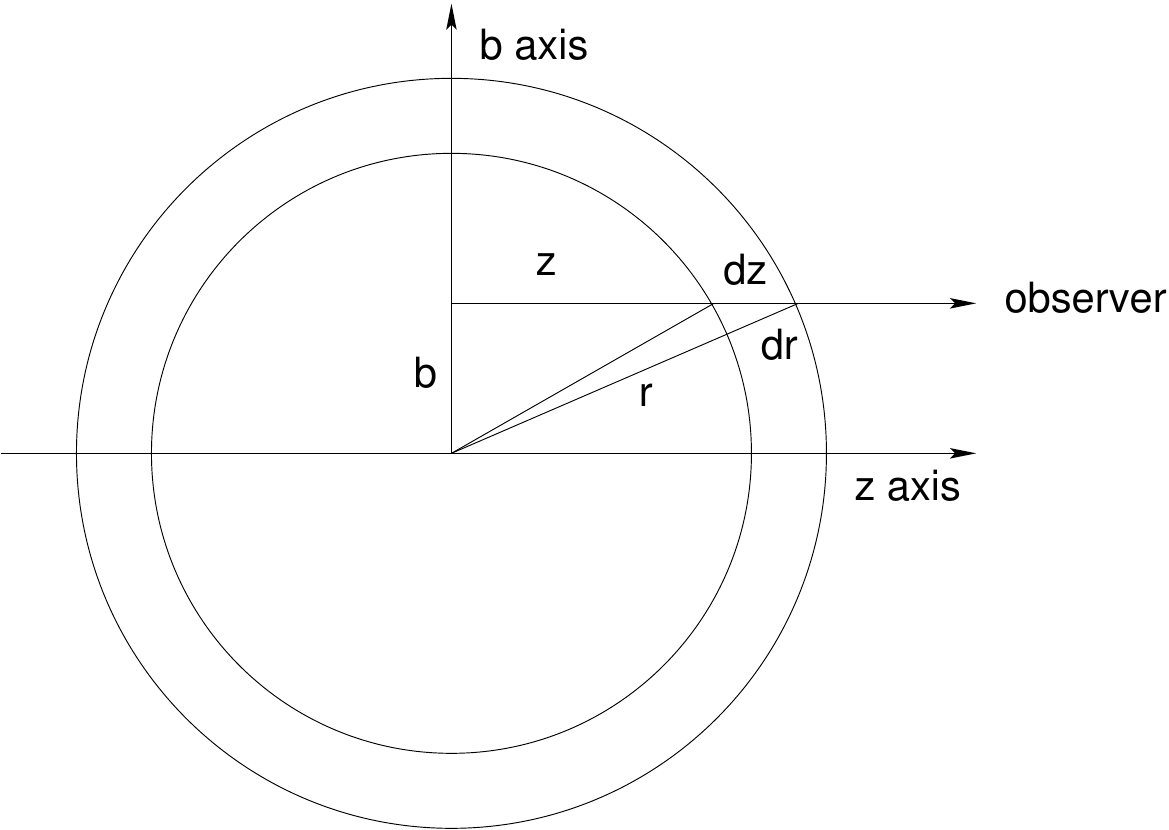}
\caption{\label{fig:geom}
Geometry of a spherically-symmetric envelope.}
\end{figure}

Let us assume that the emission is optically thin, that the Rayleigh-Jeans approximation is valid, that $\Tex\gg\Tbg$, and that $\Trot\simeq\Tk$.
In this case, the intensities of the $(1,1)$ and $(2,2)$ main lines are given by
\begin{equation}\label{eq:Tlthin}
\Tl(j,k;\mm)= \Tex\,\tau(j,k;\mm).
\end{equation}
Equation \ref{eq:Njk} giving the column density as a function of the optical depth, in the optically thin case, can be simplified to
\begin{equation}\label{eq:Nthin}
N_{jk}= 
\sqrt{\frac{\pi}{4\ln2}} \,
\frac{16\pi k\nu_{jk}^2}{hc^3A_{jk}R_{jk\mm}}\, 
\Tex \, \tau(j,k;\mm) \, \Delta v,
\end{equation}
so that, in the optically thin case, the column densities are proportional to the line intensities, and their ratio can be obtained directly from the ratio of line intensities,
\begin{equation}\label{eq:Nratiothin}
\frac{N_{11}}{N_{22}} = \frac{A_{22}R_{22\mm}}{A_{11}R_{11\mm}} \frac{\Tl(1,1;\mm)}{\Tl(2,2;\mm)}.
\end{equation}
Thus, the homogeneous temperature can be simply derived from the ratio of intensities of the $(1,1)$ and $(2,2)$ lines. From Eqs.\  \ref{eq:Nratiothin} and \ref{eq:trot}, we have
\begin{equation}\label{eq:trotthin}
\left[\frac{\Th}{\mathrm{K}}\right]=
\frac{40.99}
{\displaystyle\ln\left(3.544\frac{T(1,1;\mm)}{T(2,2;\mm)}\right)}.
\end{equation}

Let us estimate, for a given projected radius $b$, the column densities $N_{11}$ and $N_{22}$,
\begin{equation}
N_{jk}(b) = \int\sub{line~of~sight} n_{jk}(r) \, dz,
\end{equation}
where $dz$ is the elementary length along the line of sight, and $r=\sqrt{b^2+z^2}$ (see Fig.\  \ref{fig:geom}). 
The function that is integrated, the number density $n_{jk}(r)$, depends on 
the fractional abundance $f_{jk}$, which depends on the envelope temperature (Eq.\ \ref{eq:f11f22}), and the envelope density, $n(r)$,
\begin{equation}
n_{jk}(r)= f_{jk}\left(T(r)\right)\,X\,n(r).
\end{equation}
If we assume that the density is given by a simple power law, the maximum of $n_{jk}$ occurs at the minimum value of $r$, $r\sub{min}=b$, with a value $n_{jk}(b)$, and we can roughly approximate the integral along the line of sight by
\begin{equation}
\int\sub{line~of~sight}\!\!\!\! n_{jk}(r) \, dz \simeq n_{jk}(b)\,2b.
\end{equation}
Thus, with this approximation, the column density ratio for a projected radius $b$ equals the fractional density radio for $r=b$,
\begin{equation}
\frac{N_{11}(b)}{N_{22}(b)}= \frac{n_{11}(r=b)}{n_{22}(r=b)}.
\end{equation}
Since the envelope temperature gives the fractional density ratio $f_{11}/f_{22}$ (Eq.\  \ref{eq:f22}), and the same expression for the column density ratio gives the homogeneous temperature (Eq.\  \ref{eq:trot}), both temperatures coincide.

In conclusion, in the optically thin case and with the rough approximation made for the integral along the line of sight, and for a power-law density, the homogeneous temperature coincides with the envelope temperature,
\begin{equation}
\Th(b) = T(b) = T_0 \, (b/r_0)^{-q}.
\end{equation}
However, the \nh{} $(1,1)$ emission is in many cases partially thick, and the optically thin approximation does not hold. 
In addition, even for a power-law density, the maximum \nh{} $(1,1)$ volume density $n_{11}$ is not necessarily found at the minimum distance of the line of sight to the envelope center, because for temperatures greater than $\sim20$ K, the density  $n_{11}$ decreases with temperature (Fig.\  \ref{fig:njk_t}), and the approximation used to estimate the column density does not hold.

\section{Radiative transfer calculation}
\label{sec:model}

A radiative transfer code was developed to derive the intensities of the \nh{} $(1,1;\mm)$, $(1,1;\ms)$, and $(2,2;\mm)$ lines for a spherical envelope, and calculate the homogeneous optical depths $\tau\sub{hom}(1,1;\ms)$ and $\tau\sub{hom}(2,2;\ms)$, and the homogeneous temperature $\Th$ derived from the homogeneous analysis. The description of the method is given in Appendix \ref{appendix:model}.

\begin{figure}
\centering
\includegraphics[width=0.85\columnwidth]{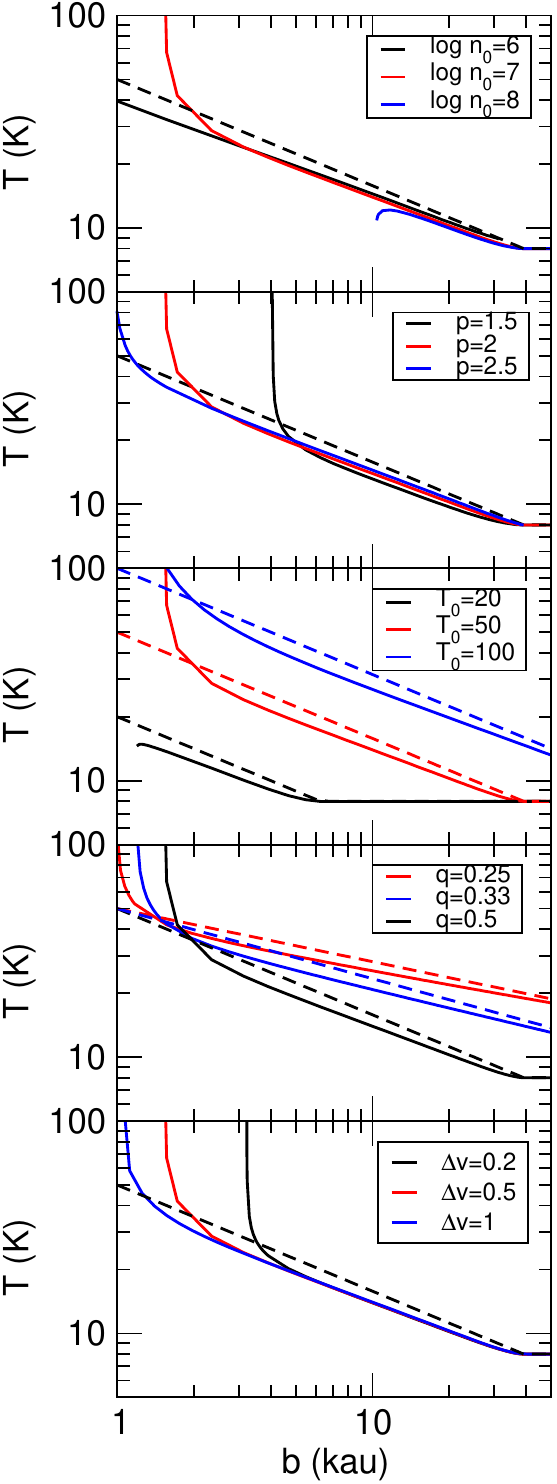}
\caption{
Radial profiles  of the homogeneous temperature $\Th$ (solid lines) for different parameters of an envelope with a power-law density profile.
The dashed lines indicate the radial profiles of the actual kinetic temperature of the envelope.
For all panels, except for the parameter indicated in each panel, the envelope is at a distance of 1 kpc, with 
$n_0= 10^7$ \cmt, $p= 2$, $r_c=0$,
$T_0= 50$ K, $q= 0.5$, and
$\Delta v= 0.5$ \kms.
\textit{Top panel:} Envelope densities $n_0= 10^6$, $10^7$, and $10^8$ \cmt{}.
\textit{Second panel:} Density power-law indices $p= 1.5$, 2, and 2.5.
\textit{Third panel:} Envelope temperatures $T_0= 20$, 50, and 100 K.
\textit{Fourth column:} Temperature power-law indices $q= 0.25$, 0.33, and 0.5.
\textit{Bottom column:} Line widths $\Delta v= 0.2$, 0.5, and 1 \kms{}. 
}
\label{fig:thom_all}
\end{figure}

\begin{figure}
\centering
\includegraphics[width=0.8\columnwidth]{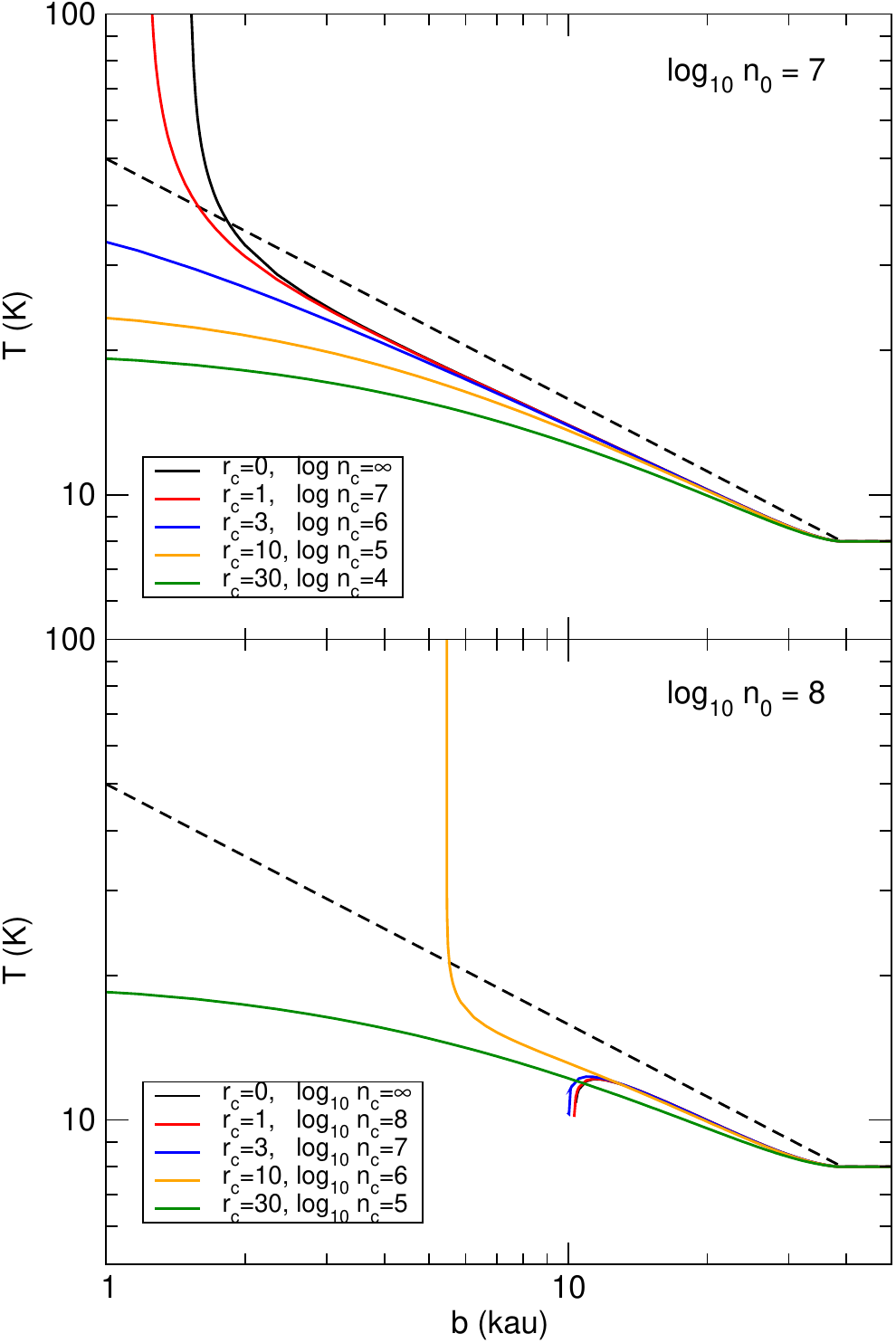}
\caption{
Radial profiles  of the homogeneous temperature $\Th$ (solid lines) for an envelope with a Plummer-like density profile with different values of the radius of the inner flat region, $r_c= 0$, 1, 3, 10, and 30 kau.
The top panel is for a density $n_0=10^7$ \cmt{}, and the bottom panel for $n_0=10^8$ \cmt{}.
The dashed lines indicate the radial profiles of the actual kinetic temperature of the envelope.
For both panels the envelope is at a distance of 1 kpc, with 
$p= 2$,
$T_0= 50$ K, $q= 0.5$, and
$\Delta v= 0.5$ \kms.
The values of $r_c$ (kau) and the resulting central densities $n_c$ (\cmt{}) are indicated in each panel.
In the bottom panel, the profiles for $r_c=0$, 1, and 3 are nearly coincident.
}
\label{fig:thom_allrc}
\end{figure}

\subsection{Results}
\label{sec:results}

A summary of results for different envelope parameters is shown in Figs.\ \ref{fig:thom_all} and \ref{fig:thom_allrc}.
In Fig.\ \ref{fig:thom_all} we show the actual temperature profile and the homogeneous temperature profile obtained for envelopes with power-law density profiles ($r_c=0$), and in Fig.\ \ref{fig:thom_allrc} with Plummer-like density profiles ($r_c > 0$). 
The parameters of the envelopes, at a distance of 1 kpc, are the same for all the panels of the figures, unless stated otherwise,
reference radius $r_0= 1$ kau,
density parameter $n_0=10^7$ \cmt{}, 
density power-law index $p= 2$, 
temperature at $r_0$, $T_0= 50$ K, 
temperature power-law index $q= 0.5$,
line width $\Delta v= 0.5$ \kms{},
minimum density $n\sub{min}=10^3$ \cmt{},
minimum temperature $T\sub{min}=8$ K, 
and \nh{} abundance $X= 10^{-8}$.
For each panel of Figs.\ \ref{fig:thom_all} and \ref{fig:thom_allrc} a single parameter of the envelope, indicated in the panel, is changed.

In Fig.\  \ref{fig:thom_all} (power-law density, $r_c= 0$),
the first panel shows the results for different densities,
$n_0= 10^6$ \cmt{} (resulting $r\sub{env}=0.15$ pc), 
$10^7$ \cmt{} ($r\sub{env}=0.5$ pc) and 
$10^8$ \cmt{} ($r\sub{env}=1.5$ pc). 
The second panel shows different values of the density power-law index, $p=1.5$ (resulting $r\sub{env}=2.2$ pc), 
$p=2$ ($r\sub{env}=0.5$ pc) and 
$p=2.5$ ($r\sub{env}=0.08$ pc). 
The third panel shows different values of the envelope temperature, 
$T_0= 20$, 50, and 100 K.
The fourth panel, different values of the temperature power-law index, $q=0.25$, 0.33, and 0.5.
Finally, the bottom panels shows different values of the line width, $\Delta v= 0.2$, 0.5, 1 \kms{}.
For the three bottom panels the resulting envelope radius is the same, $r\sub{env}=0.5$ pc.

In Fig.\ \ref{fig:thom_allrc} we show the results for Plummer-like density profiles for different values of the radius of the inner flat region, $r_c=0$,
1, 3, 10, and 30 kau.
The top panel is for a density parameter $n_0=10^7$ \cmt{}, giving an envelope radius $r\sub{env}\simeq0.5$ pc, and the bottom panel is for $n_0=10^8$ \cmt{}, corresponding to $r\sub{env}\simeq1.5$ pc.
The central densities, $n_c$, are indicated for each value of $r_c$.

As can be seen in Figs.\  \ref{fig:thom_all} and \ref{fig:thom_allrc}, in some cases for most sets of parameters there is a critical value of $b$ below which no value of \Th{} can be calculated. 
This happens when for small projected radii the lines become optically thick, and some of the conditions of Eqs.\  \ref{eq:cond_tau1}, \ref{eq:cond_Trot}, and \ref{eq:cond_rtau} are not fulfilled, making impossible the calculation of  \Th{}, or of both $\tau(2,2;\mm)$ and \Th{}, or even of all three parameters $\tau(1,1;\mm)$, $\tau(2,2;\mm)$, and \Th{}. 
Near the critical value \Th{} increases sharply with decreasing projected radius.
This behaviour, as explained above, is an optical depth effect. 
For small projected radii the optical depth of the lines is high. 
However, the $(2,2)$ line is optically thinner than the $(1,1)$, and thus it has more contribution from the the inner, hotter part of the envelope. 
This increase of the intensity of the $(2,2)$ line is interpreted by the homogeneous analysis as a higher \Th. 

Far from the critical value, and for projected radii larger that the radius of the inner flat region, the homogeneous temperature is a fair estimate of the envelope radial profile of temperature, although it systematically underestimates the envelope temperature.
A possible explanation is that for optically thin $(1,1)$ and $(2,2)$ lines, the material along the line of sight contributes to the $(2,2)$ emission only if its temperature is high enough, which occurs near the center of the envelope, which is not the case for the $(1,1)$ emission. This deficit of $(2,2)$ emission is interpreted by the homogeneous analysis as a lower  \Th.

However, a much more remarkable feature of the homogeneous analysis is that for Plummer-like density profiles, the radial profile of $\Th$ can be much flatter than the temperature profile of the envelope. 
This is especially noticeably for projected radii less than the radius of the flat inner region, $b\lesssim r_c$ (see Fig.\ \ref{fig:thom_allrc}). 
It appears that the flattening of the density profile is erroneously interpreted by the homogeneous analysis as a flattening of the temperature profile.

\begin{figure}
\includegraphics[width=\columnwidth]{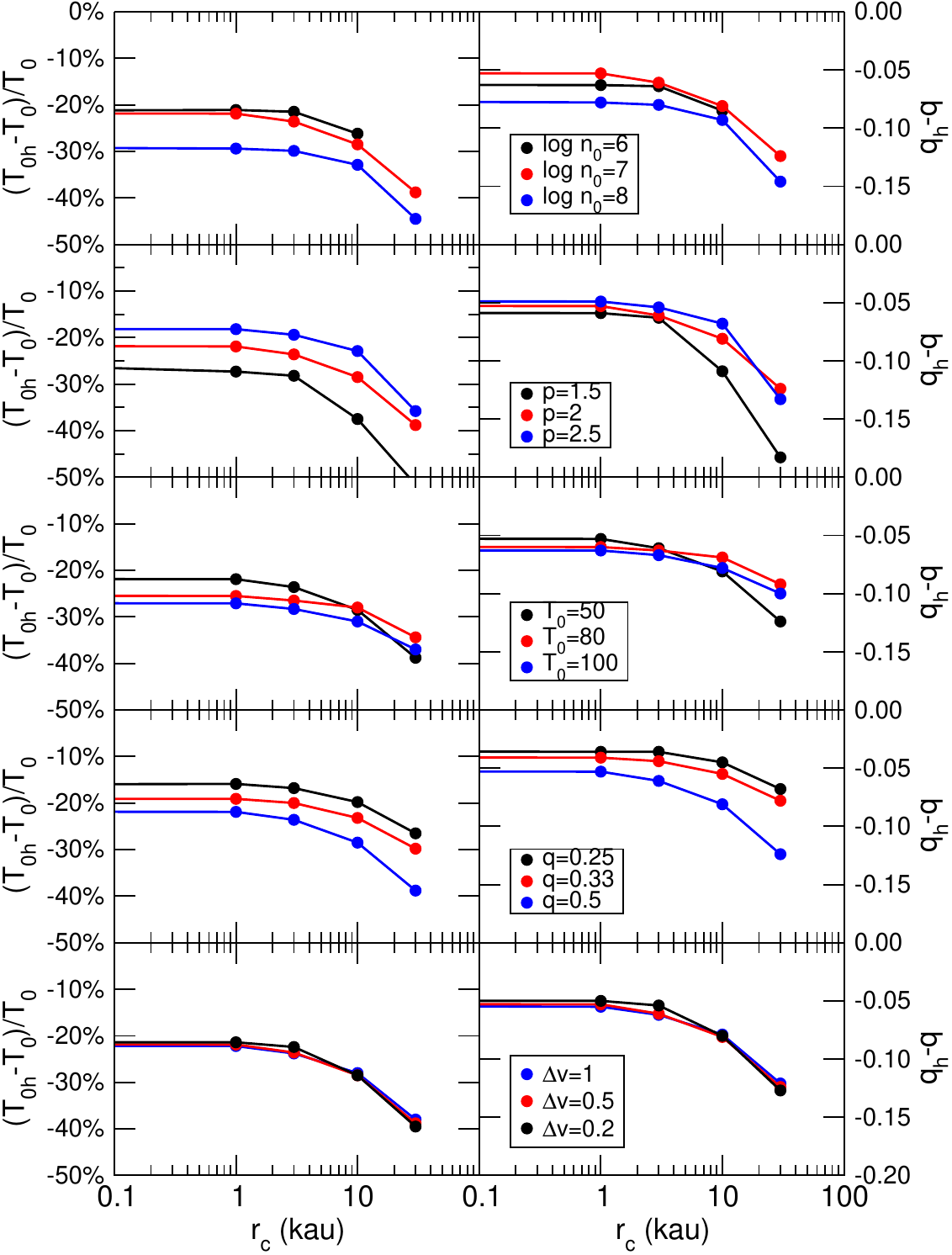}
\caption{
Comparison of the power-law radial profile of the envelope temperature, $T=T_0\,(r/r_0)^{-q}$, and of \Th{} obtained from the homogeneous analysis, for the projected radii for which it can be fitted by a power law, $\Th\simeq T\sub{0h}\,(b/r_0)^{-q\sub{h}}$.
The results obtained are shown as a function of the radius of the inner flat region $r_c$.
Each rows has two panels. 
The left panel shows the normalized difference of temperatures, $(T\sub{0h}-T_0)/T_0$, and the right panel, the difference in power-law indices, $q\sub{h}-q$. 
For each row, the three lines (black, red, and blue) correspond to different values of the parameter indicated in the right panel of the row.
The rest of parameters are the same as those used in Figs.\ \ref{fig:thom_all} and \ref{fig:thom_allrc}.
}
\label{fig:panel}
\end{figure}

In order to see the effect of the radius of the inner flat region and the rest of envelope parameters on the difference between $\Th$ and the envelope temperature, we calculated for several values of $r_c$ and different physical parameters of the envelope the radial profile of \Th{} as a function of $r_c$, and fitted a power law for the range of projected radii where the profile could be approximated well by a power law, 
\begin{equation}
\Th\simeq T\sub{0h}\,(b/r_0)^{-q\sub{h}},    
\end{equation}
obtaining values of $T\sub{0h}$ and $q_\mathrm{h}$. 
These were compared with the values of $T_0$ and $q$ of the envelope.
The results are shown in Fig.\  \ref{fig:panel}.
The results are shown as a function of $r_c$, ranging from 0 to 30 kau.
The ranges of the physical parameters of the envelope examined were
$10^6 \hbox{ \cmt{}} \le n_0 \le 10^8$ \cmt{} (top row of Fig.\  \ref{fig:panel}),
$1.5 \le p \le 2.5$ (second row),
$50\mathrm{~K} \le T \le 100$ K (third row), 
$0.25 \le q \le 0.5$ (fourth column), and
$0.2 \le \Delta v \le 1$ \kms{} (bottom row).
These ranges include the typical values derived for massive dense cores and clumps in previous works \citep[e.g.,][]{Beu02,Will05,Gia13,Pal14,Pal21,Gie21}.
Two panels are shown for each row, the left panel with the normalized difference of temperatures, $(T\sub{0h}-T_0)/T_0$, and the right panel with the power-law indices difference, $q\sub{h}-q$.
For each row the three lines and dots correspond to different values of the parameter indicated in the same row.
As can be seen in the figure, for small values of $r_c$ the difference $(T\sub{0h}-T_0)/T_0$ is typically $-20$\% to $-30$\%, and $q\sub{h}-q$ of the order of $-0.05$.
However, these differences increase dramatically for higher values of the inner flat region. 
The temperature difference can be as high as $-40$\% to $-50$\%, and the power-law index difference can reach $-0.10$ or $-0.15$.
The differences also increase with increasing density parameter $n_0$, density power-law index $p$, and temperature power-law index $q$.

\subsection{Effect of beam smoothing}

\begin{figure}
\centering
\includegraphics[width=0.85\columnwidth]{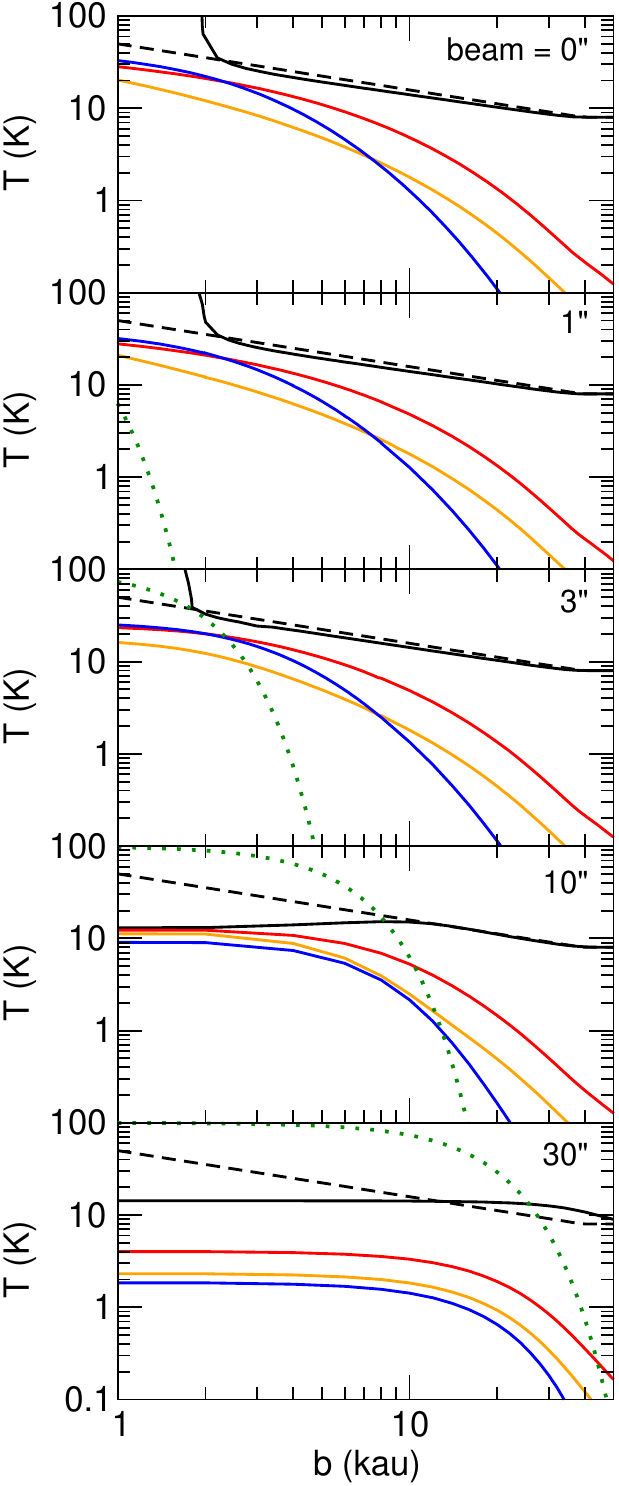}
\caption{
Effect of beam smoothing on the radial profiles of the
line intensities
$\Tl(1,1;\mm)$ (red),  
$\Tl(1,1;\ms)$ (orange),  
$\Tl(2,2;\mm)$ (blue), 
and homogeneous temperature $\Th$ (black continuum line).
The temperature of the envelope is indicated with a black dashed line.
The beam HPBW (in arcsec) is indicated in each panel, and the beam profiles, with an arbitrary height at the origin of 100, are shown as green dotted lines.
The parameters of the envelope are
$n_0= 10^7$ \cmt, $p= 2$,
$T_0= 50$ K, $q= 0.5$, and
$\Delta v= 0.5$ \kms{}
The distance is 1 kpc. 
At this distance, 1 arcsec corresponds to a length of 1 kau.
}
\label{fig:thom_beam}
\end{figure}

\begin{figure}
\centering
\includegraphics[width=0.85\columnwidth]{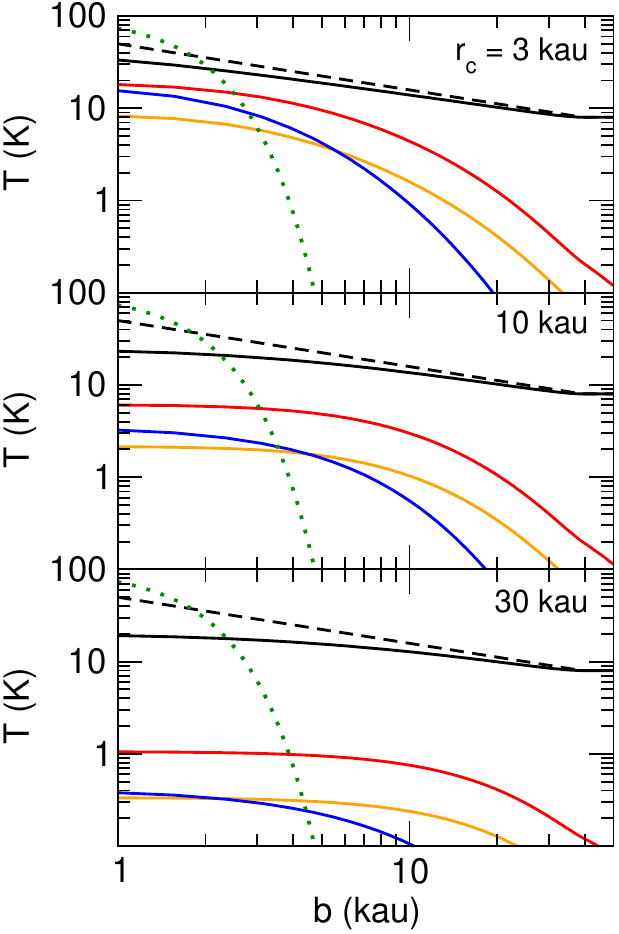}
\caption{
Effect of beam smoothing for a Plummer-like density profile for different values of the radius of te inner flat region, $r_c= 0.3$, 1, and 3 kau. 
The radial profiles of the line intensities are shown as 
a red line for $\Tl(1,1;\mm)$,  
an orange line for $\Tl(1,1;\ms)$, and 
a blue line for $\Tl(2,2;\mm)$.
The homogeneous temperature $\Th$ is shown as a black solid line, and
the temperature of the envelope is indicated with a black dashed line.
The beam HPBW  is $5''$ for all panels and the beam profile, with an arbitrary height at the origin of 100, is shown as a green dotted line.
The parameters of the envelope are
$n_0= 10^7$ \cmt, $p= 2$,
$T_0= 50$ K, $q= 0.5$, and
$\Delta v= 0.5$ \kms{}
The distance is 1 kpc. 
At this distance, 1 arcsec corresponds to a length of 1 kau.
}
\label{fig:thom_beam_rc}
\end{figure}

Up to now we have not considered the effect of beam smoothing on the observed profiles.
Beam smoothing will produce a flattening of the observed radial profiles at the center of the envelope, in a region with a size of the order of the beam size.

In order to model the effect of a finite beam size, the radial profiles of the line intensities $\Tl(1,1;\mm)$, $\Tl(1,1;\ms)$, and $\Tl(2,2;\mm)$ were converted to 2D-maps. 
The maps were then 2D-convolved with a Gaussian beam pattern, and the resulting maps were ring-averaged to obtain the beam-smoothed radial profiles of the line intensities. 
From these, the homogeneous analysis was performed to obtain the radial profile of homogeneous temperature \Th{}.

In Fig.\ \ref{fig:thom_beam} we show the results for an envelope with a power-law density ($r_c=0$), at a distance of 1 kpc, for different beam sizes: no beam smoothing (top panel), and half-power beam widths (HPBW) of $1''$, $3''$, $10''$, and $30''$ (second to fifth panels). 
As can be seen in the figure, the effect of beam smoothing is not noticeable for beam sizes less than $3''$. 
The reason is that for the parameters of density and temperature of the envelope, the homogeneous temperature could not be derived for projected radii less than $\sim2''$, where the effect of beam smoothing is important.
However, the effect of beam smoothing is very apparent for beam sizes of $10''$ and $30''$, for which the radial profile of \Th{} becomes flat for projected radii less than approximately the HPBW.

The result of beam smoothing for the same envelope, but with a Plummer-like density profile is shown in Fig.\ \ref{fig:thom_beam_rc}.
In all three panels the beam is the same, $\mbox{HPBW}=3''$, but the radius of the inner flat region is different: $r_c= 3$ kau (top panel), 10 kau (center), and 30 kau (bottom).
As can be seen, the discrepancy between the homogeneous temperature and the envelope temperature becomes important for values of $r_c$ larger than the beam size.

\section{Application to G14.225$-$0.506}

\subsection{G14.2 Hub N and S}

A well-suited observational case to apply the study presented above is the infrared dark cloud G14.225$-$0.506 (hereafter G14.2), located at a distance of 1.98 kpc \citep{Xu11}. 
The cloud consists of two main hubs, called Hub N and Hub S, being the center of a number of filamentary structures. 
This cloud has been deeply studied in NH$_3$(1,1) and (2,2) using a combination of single-dish and interferometer. 
In particular, the cloud has been observed in NH$_3$(1,1) and (2,2) with high angular resolution and with no missing short spacings, allowing to perform a reasonable study of its structure. 
A detailed description of the  main observational results on G14.2 can be found in \citet{Bus13}, \citet{Bus16}, \citet{San16}, \citet{Oha16}, and \citet{Che19}.

\subsection{G14.2 data description}

The \nh{} $(J,K)=(1,1)$ and $(2,2)$ observations were carried out with the Very Large Array and the Effelsberg radio telescope. 
The data were combined, resulting in a final synthesized beam of $8\farcs17\times7\farcs00$, $\mbox{P.A.}=-15\fdg1$, and maps with a pixel size of $2''$. 
A full description of the observations and data reduction can be found in \citet{Bus13}.

\begin{figure}
\centering
\includegraphics[width=\columnwidth]{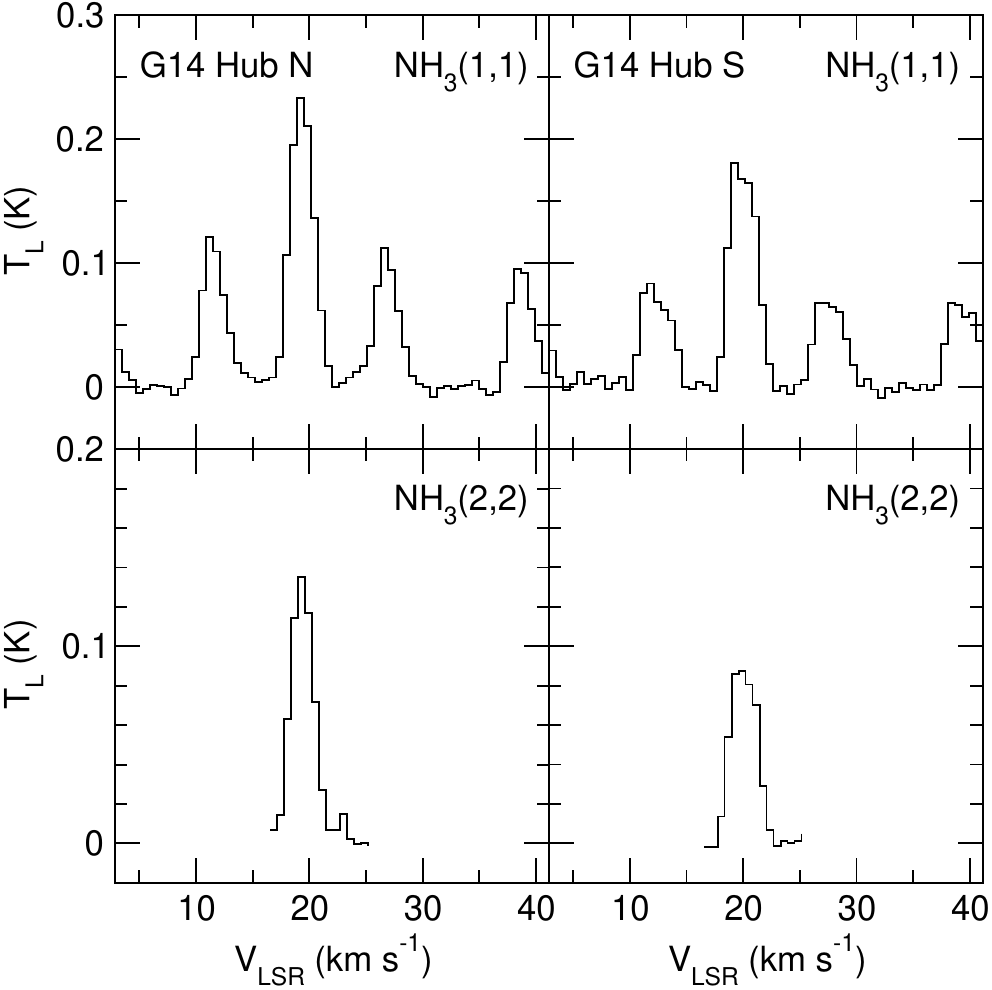}
\caption{
G14.2: \nh{} $(1,1)$ (top) and $(2,2)$ (bottom) spectra at the position of the center of Hub N (left) and Hub S (right), averaged within a diameter of $4''$.
}
\label{fig:g14_spectra}
\end{figure}

The $(1,1)$ line was observed with 63 channels 0.62 \kms{} wide, with a velocity range between 2.8 and 41.2 \kms{}, encompassing the main line, the inner satellite lines, and one of the outer satellite lines.
For the $(2,2)$ line, the number of channels was 15, with the same spectral resolution, and a velocity range between 16.5 and 25.2 \kms{}, encompassing the main line (see Fig.\  \ref{fig:g14_spectra}).

\subsection{G14.2 data analysis}
\label{sec:pixel_fit}

\begin{figure}
\centering
\includegraphics[width=0.91\columnwidth]{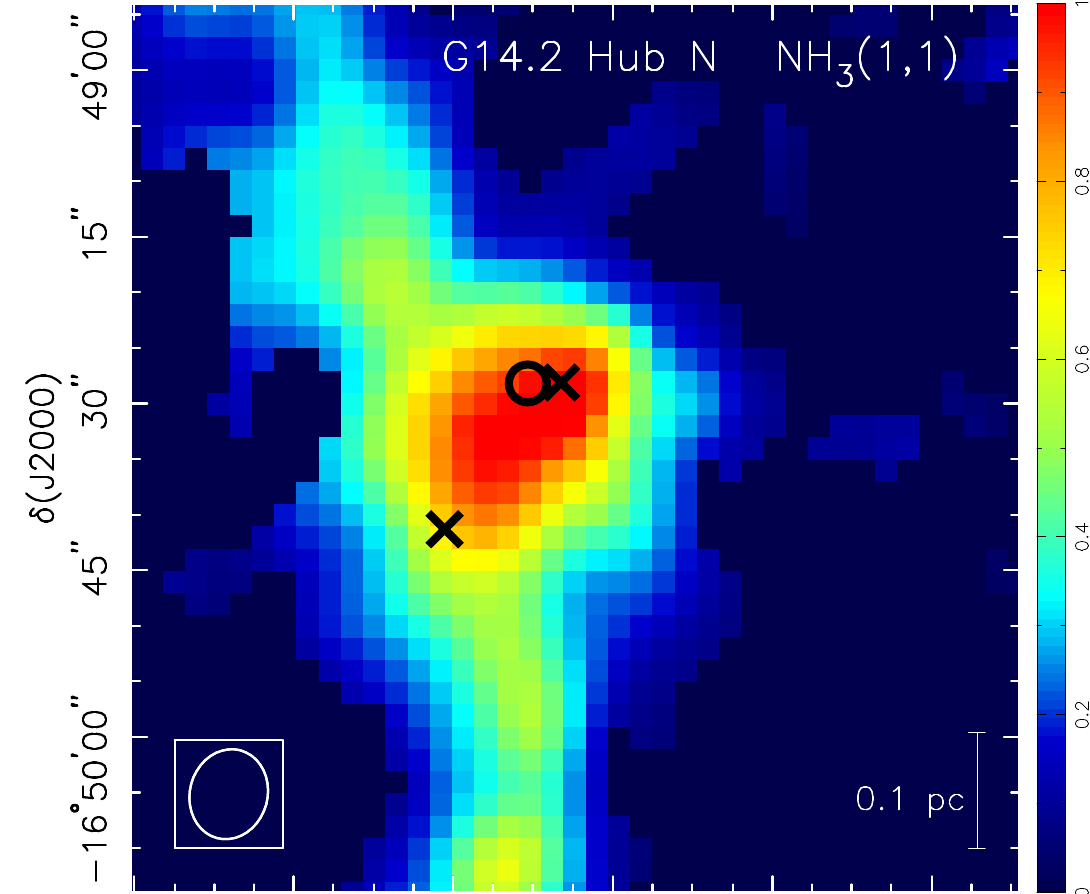}
\includegraphics[width=0.91\columnwidth]{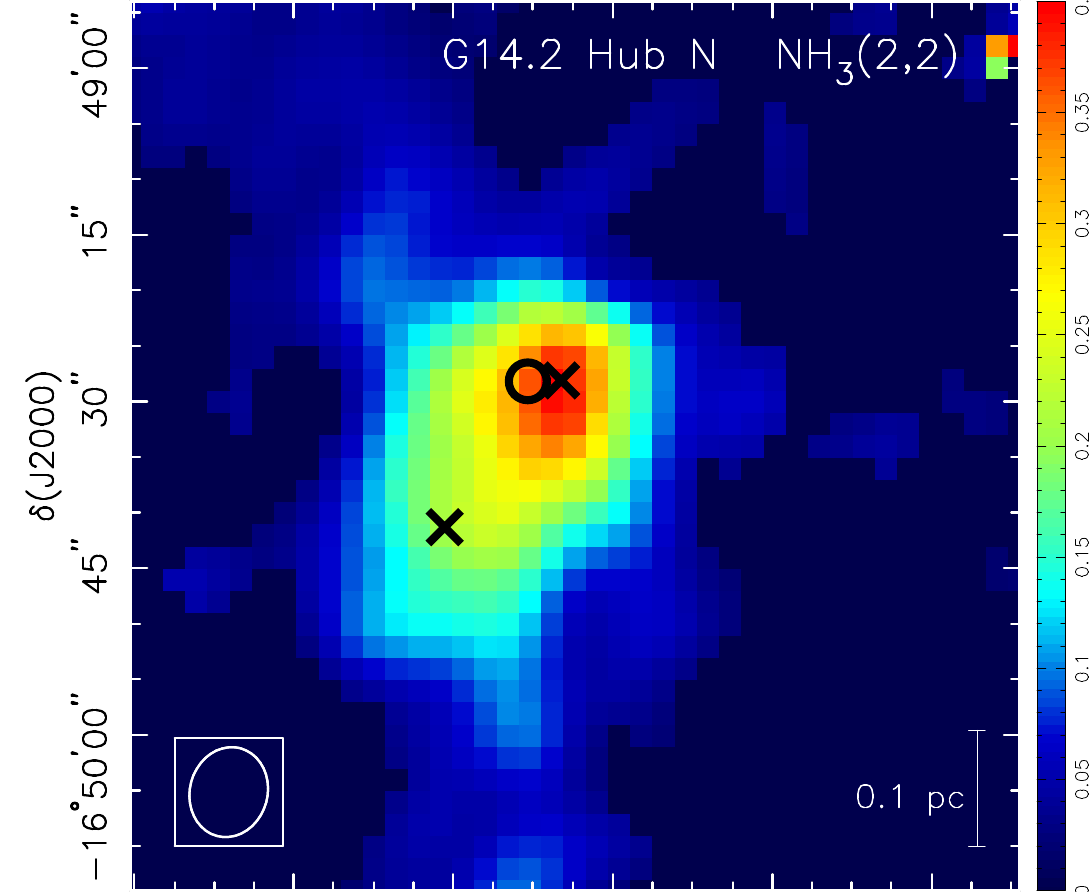}
\includegraphics[width=0.91\columnwidth]{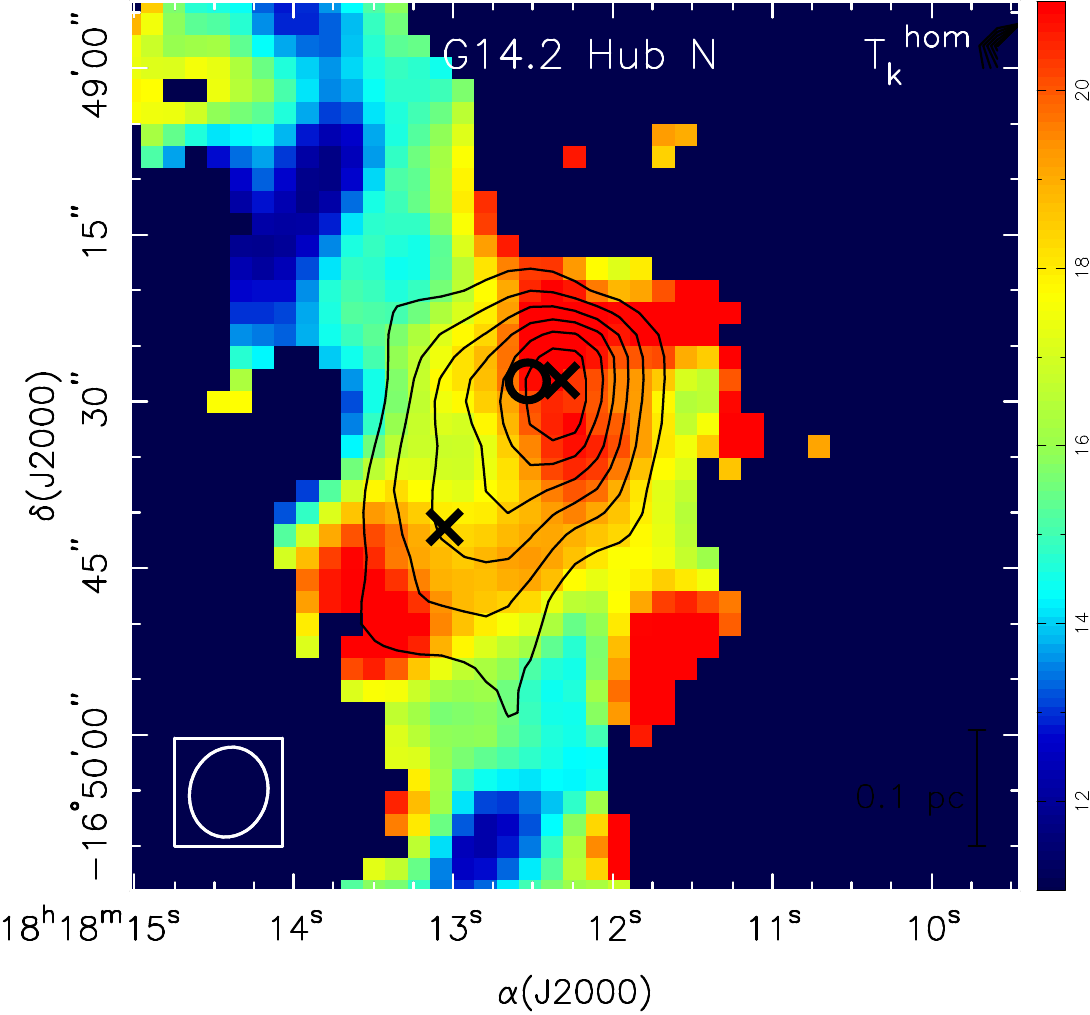}
\caption{
G14.2, Hub N. 
\textit{Top and middle panels:} \nh{} $(1,1)$ and $(2,2)$ main line integrated intensities, corrected for opacity, $A\tau_{m}\Delta v$ (see main text in \S{} \ref{sec:pixel_fit}).
The color scale is in units of K \kms{}.
\textit{Bottom panel:} homogeneous temperature \Th{}.
The color scale for \Th{} goes from 10 to 21 K. The contours are the integrated \nh{} $(2,2)$ emission, starting at 0.1 K \kms{} with steps of 0.05 K \kms{}.
For all panels the crosses mark the positions of the SMA mm sources MM1 and MM3, in order of increasing RA, and the circle marks the position of the 350 $\mu$m emission peak \citep{Bus16}.
The beam is shown in the bottom-left corner of each panel.
}
\label{fig:hubn_atau1mdv}
\label{fig:hubn_atau2mdv}
\label{fig:hubn_Tk}
\end{figure}

\begin{figure}
\centering
\includegraphics[width=0.89\columnwidth]{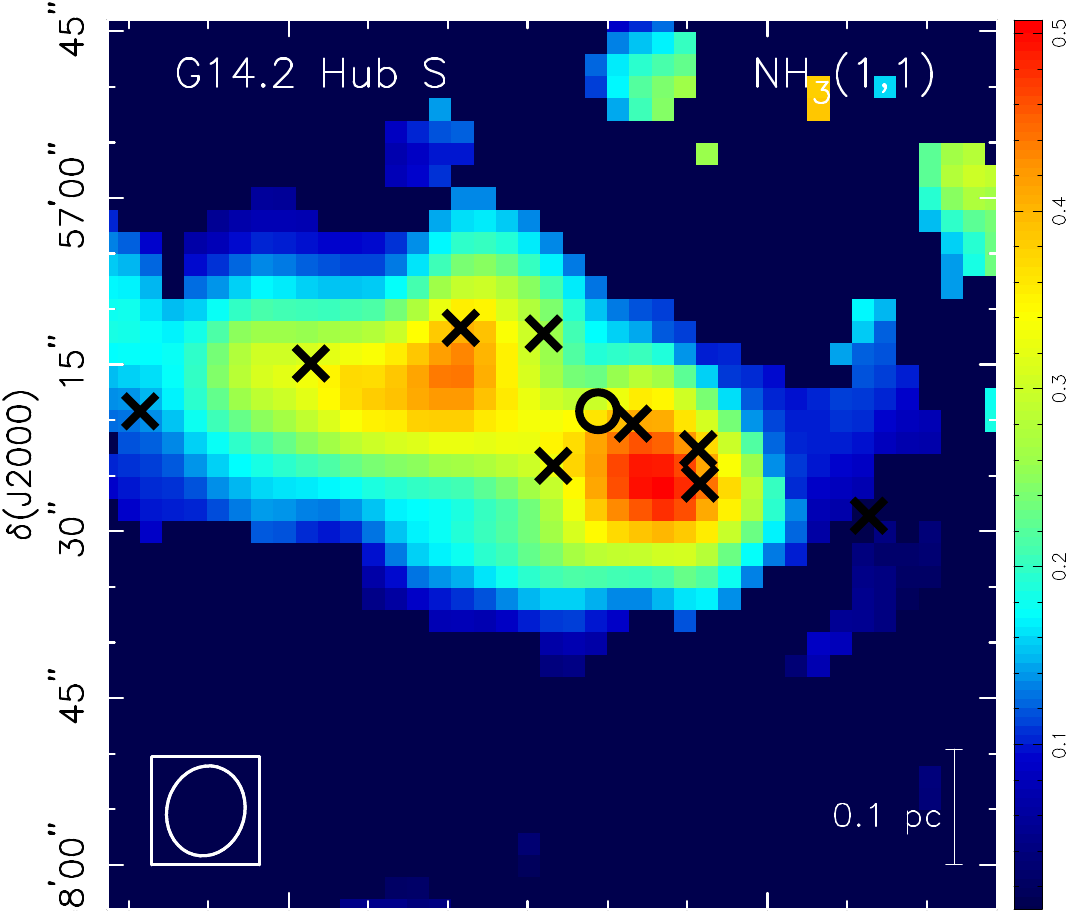}
\includegraphics[width=0.89\columnwidth]{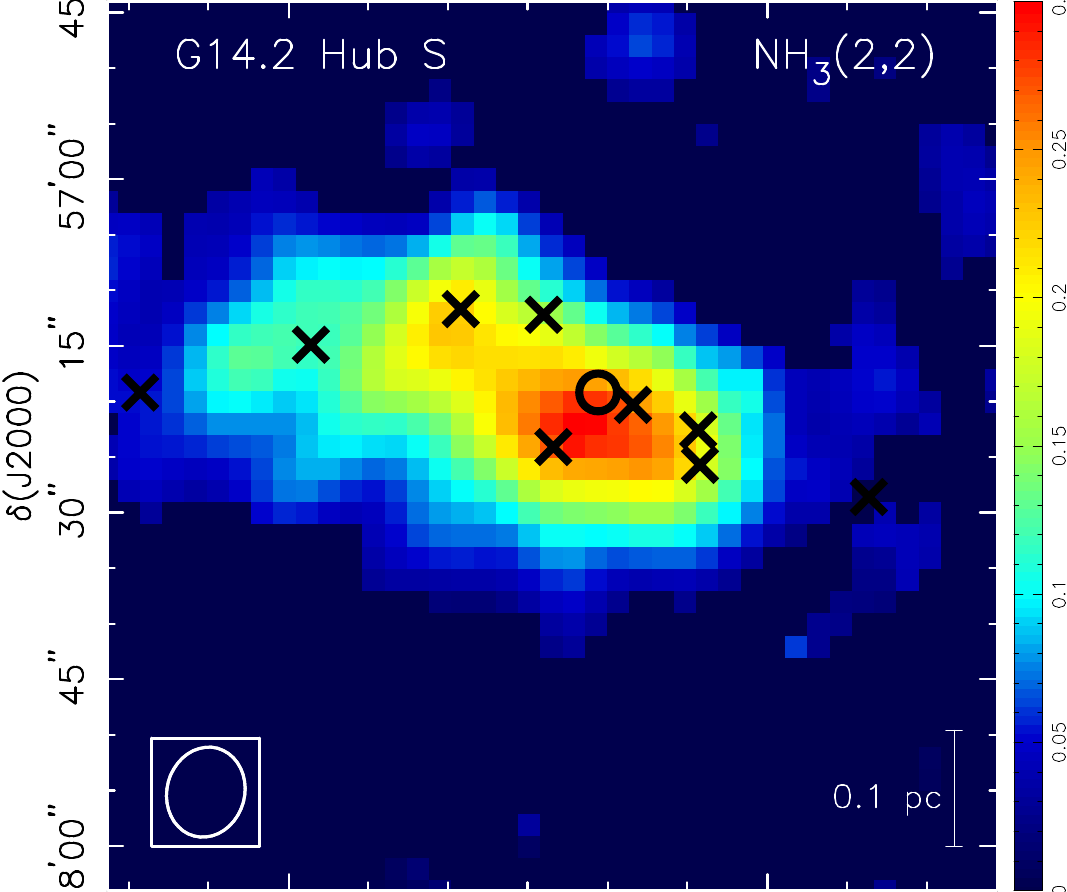}
\includegraphics[width=0.89\columnwidth]{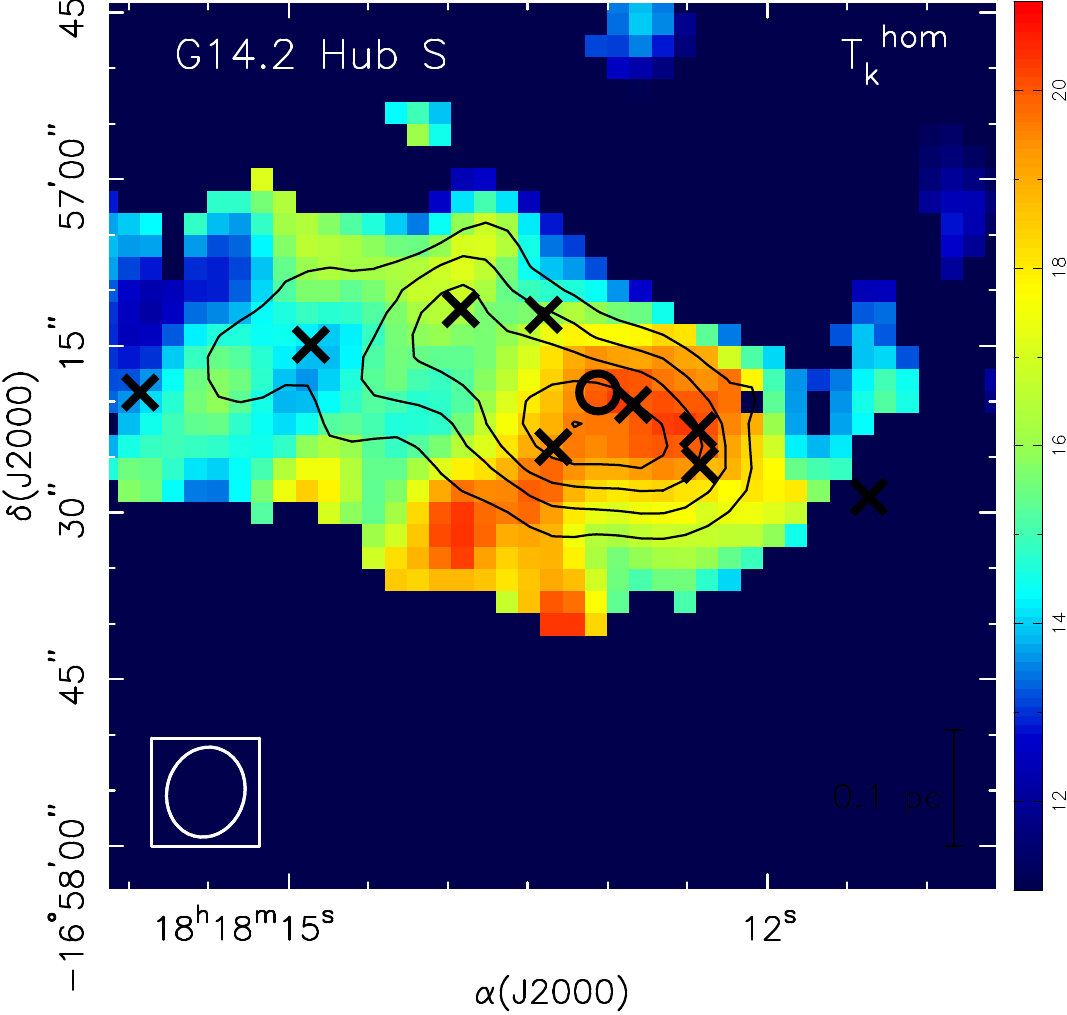}
\caption{
G14.2, Hub S.
\textit{Top and middle panels:} \nh{} $(1,1)$ and $(2,2)$ main line integrated intensities, corrected for opacity, $A\tau_{m}\Delta v$ (see main text in \S{} \ref{sec:pixel_fit}).
The color scale is in units of K \kms{}.
\textit{Bottom panel:} homogeneous temperature \Th{}.
The color scale for \Th{} goes from 10 to 21 K. The contours are the integrated \nh{} $(2,2)$ emission, starting at 0.1 K \kms{} with steps of 0.05 K \kms{}.
For all panels the crosses mark the positions of the SMA mm sources MM2 to MM9, in order of increasing right ascension, and the circle marks the position of the 350 $\mu$m emission peak \citep{Bus16}.
The beam is shown in the bottom-left corner of each panel.
}
\label{fig:hubs_atau1mdv}
\label{fig:hubs_atau2mdv}
\label{fig:hubs_Tk}
\end{figure}

\begin{figure}
\centering
\includegraphics[width=\columnwidth]{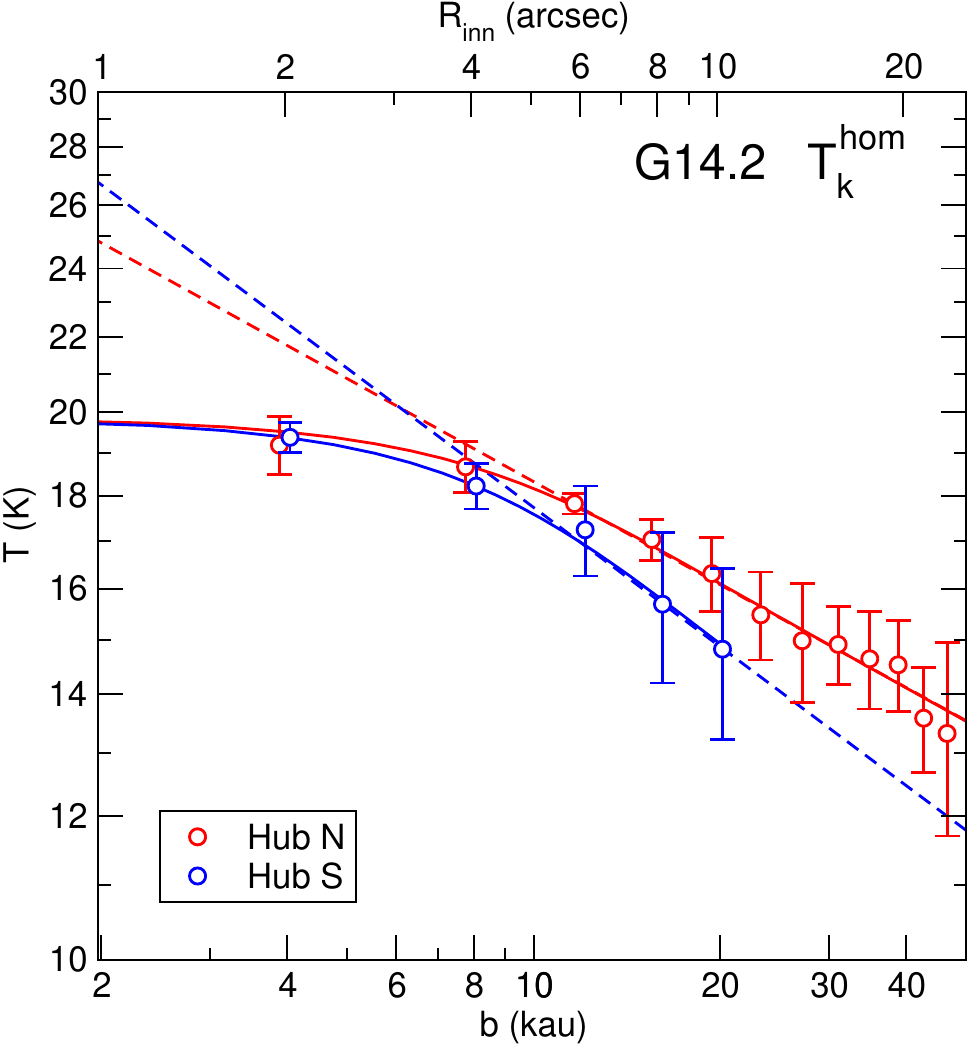}
\caption{
G14.2: \Th{} radial profile obtained from the maps of Figs.\ \ref{fig:hubn_Tk} and \ref{fig:hubs_Tk} (small circles and error bars), for Hub N (red) and Hub S (blue), in a log--log scale.
The error bars indicate the rms dispersion of the values of $\Tk$ averaged in each ring.
The red and blue solid lines are the best fit to the radial profiles, with the 2D-convolution of a Gaussian beam of $7\farcs6$ with a power law $\Th=T\sub{0h}(r/r_0)^{-q\sub{h}}$.
The dashed lines are the \Th{} power-law profiles deconvolved from the Gaussian beam.
}
\label{fig:g14_Thom_radial}
\label{fig:g14_Tk_T0fit}
\end{figure}

The \nh{} $(1,1)$ and $(2,2)$ lines in the two hubs were analyzed using the \texttt{hfs\_nh3\_cube} utility of the HfS tool \citep{Est17}. 
For every pixel, the spectra of the pixels inside a circle of $2''$ radius were averaged and the resulting $(1,1)$ and $(2,2)$ spectra were fitted simultaneously, obtaining the parameters of the lines: line width, central velocities and intensities of the $(1,1)$ and $(2,2)$ lines, and optical depths of the $(1,1)$ and $(2,2)$ main lines. 
From these, the standard homogeneous analysis provided the physical parameters pixel to pixel: excitation, rotational, kinetic temperatures, and column density of \nh{}. 

We show for Hub N (Fig.\ \ref{fig:hubn_atau1mdv}) and Hub S (Fig.\ \ref{fig:hubs_atau1mdv}), the integrated intensity, corrected for opacity, $A\tau_{m}\Delta v$, of the $(1,1)$ line (top panel) and $(2,2)$ line (middle panel), where $A$ is the amplitude $A=\eta\sub{B}[J_\nu(\Tex)-J_\nu(\Tbg)]$, $\eta\sub{B}$ is the beam filling factor, $\tau_m$ is the main line optical depth, and $\Delta v$ is the line width.
The figure also shows the positions of the millimeter dust continuum sources detected with the Submillimeter Array (SMA) by \citet{Bus16}.

The maps of the homogeneous temperature, $\Th$, for Hub N and Hub S are shown in the bottom panels of the same Figs.\  \ref{fig:hubn_Tk} and \ref{fig:hubs_Tk}. 
The \nh{} $(2,2)$ integrated intensity are also shown as contours.
These maps were used to average the values of $\Th$ 
in rings, $2''$ wide, centered on the position of the emission peaks in the hubs N and S at 350 $\mu$m observed with the Caltech Submillimeter Observatory (CSO) reported by \citet{Bus16}.
The CSO positions were chosen as center of the rings because the CSO angular resolution ($9''$) is very similar to that of the \nh{} observations, and thus both observations are sensitive to similar structures at the same scale.
These positions are shown as small circles in Figs.\ \ref{fig:hubn_Tk} and \ref{fig:hubs_Tk}. 
The positions coordinates are 
\RA{18}{18}{12}{53}, \DECf{-16}{49}{28}{2} (Hub N), and
\RA{18}{18}{13}{06}, \DECf{-16}{57}{19}{2} (Hub S).
The radial profiles obtained for $\Th$ are shown in Fig.\  \ref{fig:g14_Thom_radial} as small circles and error bars. 
The error bars are the rms dispersion of the values of $\Th$ averaged in each ring.
The maximum radius and range of position angles (PA) used are given in Table \ref{tab:g14_Trot}. 
They were chosen to avoid contamination from other dense clumps in the region, which produce an increase in temperature to the southeast of the center of the rings in both hubs.
By constraining the range of PAs, we assure that the profiles are obtained in smooth regions unaffected by feedback from nearby sources.

\begin{table}
\centering
\caption{G14.2 Hub N and S. Results from the fit of the observed \Th{} radial profiles with the 2D-convolution of a power-law $T= T\sub{0h} (r/r_0)^{-q\sub{h}}$ with a Gaussian beam of $7\farcs56$.
\label{tab:g14_Trot}
}
\begin{tabular}{lccccc}
\hline
      & ${T\sub{0h}}^a$ & & & ${R\sub{max}}^b$ & PA range$^b$ \\
G14.2 & (K) & $q\sub{h}$ & ${\chi_r}^c$ & (arcsec) & (deg) \\
\hline
Hub N & 28 & 0.19 & 0.33 & 26 & \phs20 -- 120 \\
Hub S & 31 & 0.25 & 0.14 & 12 &  $-60$ -- 120 \\
\hline
\end{tabular}

$^a$ Temperature at $r_0= 1$ kau.
$^b$ Maximum radius and range of position angles used to obtain the \Th{} radial profile.
$^c$ $\chi_r=\sqrt{\chi^2/(n-1)}$.
\end{table}

\subsection{Homogeneous analysis results}

\begin{table}
\centering

\caption{
G14.2 Hub N and S. Comparison of the power-law radial profile of temperature, $T=T_0(r/r_0)^{-q}$, from the dust emission and from the homogeneous analysis.
\label{tab:g14_T0h}
}

\begin{tabular}{lccccccc}
\hline
& \multicolumn{2}{c}{\bf From dust$^a$} 
&& \multicolumn{2}{c}{\bf From \nh$^b$} \\
\cline{2-3} \cline{5-6}
    & $T_0$ &     && $T\sub{0h}$            \\
G14.2 & (K) & $q$ && (K) & $q\sub{h}$ & $(T\sub{0h}-T_0)/T_0$ & $q\sub{h}-q$ \\
\hline
Hub N & 51 & 0.34 && 28 & 0.19 & $-45$\% & $-0.15$ \\
Hub S & 45 & 0.34 && 31 & 0.25 & $-30$\% & $-0.09$ \\
\hline
\end{tabular}

$^a$ \citet{Bus16}.
$^b$ From Table \ref{tab:g14_Trot}.
\end{table}

The \Th{} radial profiles for Hub N and S are similar (Fig.\  \ref{fig:g14_Thom_radial}), showing a flat profile for small radii, less than the beam radius, and a decline for large radii. 
We fitted the observed radial profiles with the 2D-convolution of a power law $T\sub{0h}(r/r_0)^{-q\sub{h}}$ with a Gaussian with a HPBW of $7\farcs56$.
The best-fit values of the temperature at $r_0=1$ kau, $T\sub{0h}$, and the power-law index, $q\sub{h}$, are shown in Table \ref{tab:g14_Trot}.
The power-law homogeneous temperature obtained from the fit, deconvolved from the Gaussian beam, are shown as dashed lines in Fig.\ \ref{fig:g14_Thom_radial}.
As can be seen, although the results are similar for both hubs, the temperature of Hub S appears to be a bit steeper and with lower values than that of Hub N.

\citet{Bus16} obtain the temperature radial profile from the submillimeter dust emission and spectral energy distribution (SED) fitting. 
Their results are shown in the left columns of Table \ref{tab:g14_T0h}.
As can be seen, the \nh{} homogeneous analysis underestimates significantly the temperature value and power-law index of the envelope. 
Part of the differences can be attributed to the lack of dust-gas coupling in the lower-density region of the hubs \citep[see, for instance][]{Doty97}.
Indeed, \citet{Urban09} find, for densities of $10^5$ \cmt{} and dust temperatures of 50 K, that the gas temperature can be lower by $\sim20$ K.
However, the differences are compatible with the results obtained in Section \ref{sec:model} shown in Fig.\ \ref{fig:panel}.
As reported in \citet{Bus16}, both hubs have a flat inner region with a radius $r_c\simeq 20$ kau, and for such a density distribution, we can expect $T_0$ differences of the order of $-40$\%, and power-law index differences of up to $\sim -0.15$, in accordance with the results obtained from the comparison with dust emission.

\section{Discussion and Conclusions}

The spherical envelope model, using a radiative transfer calculation of the intensities of the \nh{} $(1,1;\mm)$, $(1,1;\ms)$, and $(2,2;\mm)$ lines, allowed us to compare the kinetic temperature derived from the standard \nh{} analysis, the homogeneous temperature \Th{}, with the actual temperature of the envelope.

Tests were performed with different characteristics of the envelope, similar to the typical values derived for massive dense cores and clumps.
In many cases we found that there is a critical value of the projected radius below which no value of \Th{} could be calculated, when the lines become  optically thick enough. 
Near the critical value, the homogeneous temperature can be much higher than the envelope temperature.
For larger projected radii, far from the critical value, the homogeneous temperature found from the standard \nh{} analysis was similar to the envelope temperature, but 
systematically underestimates the actual temperature $T_0$ and temperature power-law index $q$.
In the case of a power-law density, $n\propto r^{-p}$, $T_0$ is underestimated by $\sim20$\%, and $q$ by $\sim 0.05$.
However, for a density with a central flat region, that is, a Plummer-like density, $n\propto (r_c^2+r^2)^{-p/2}$, the discrepancy between the homogeneous temperature and the actual envelope temperature can be much higher, up to $\sim40$\% in $T_0$ and $\sim 0.15$ in $q$, for $r_c\simeq 10$ kau.

We applied this study to the infrared dark cloud G14.2.
We used the \nh{} $(J,K)= (1,1)$ and $(2,2)$ data obtained with the Very Large Array and the Effelsberg radio telescope for Hub N and Hub S.
The data were analyzed using the homogeneous analysis, that is, standard \nh{} analysis with the assumption of homogeneity along the line of sight, providing a radial profile of homogeneous temperature, \Th{}.
The radial profiles were fitted by the convolution of a power law $T\sub{h0}(r/r_0)^{-q\sub{h}}$ with a Gaussian beam of $7\farcs56$.
The homogeneous temperatures obtained from the G14.2 data were compared with the dust temperatures of \citet{Bus16}.
The lack of dust-gas coupling in the lower-density parts of the hubs could contribute to the observed discrepancies ($\sim40$\% in the temperature $T_0$ and $\sim0.1$ in the temperature power-law index $q$), but the discrepancies are in accordance with the expected results for envelopes with a central flat region.

Several conclusions can be drawn from this work.
Regarding the homogeneous analysis of \nh{} $(1,1)$ and $(2,2)$ radial profiles, two results are relevant.
Firstly, for small projected radii, where the optical depth of the lines is high, the homogeneous temperature can be much higher then the actual envelope temperature. 
Secondly, for larger projected radii, the temperature values and power-law indices derived from the homogeneous analysis could be seriously underestimated, depending on the density profile of the envelope.
For power-law density profiles, with a very high density at small radii, 
characteristic of very young, Class 0 objects, 
the temperature can be underestimated by $\sim20$\%, and the power-law index by $\sim0.05$.
However, for density profiles with a flat central region, 
characteristic of more evolved, Class I and beyond objects, 
the effect is more severe. 
The temperature can be underestimated by as much as $\sim40$\%, and the power-law index by $\sim0.15$.

A similar study for other commonly used dense gas tracers such as H$_2$CO and CH$_3$OCHO should be carried out to properly assess the validity of the temperature radial profiles inferred from dense gas thermometers.

\section*{Acknowledgements}

We thank the anonymous referee for his/her comments, helping to improve the paper.
R. E. and G. B. acknowledge financial support from the grants
PID2020-117710GB-I00 and CEX2019-000918-M funded by MCIN/ AEI /10.13039/501100011033.
A. P. acknowledges financial support from the UNAM-PAPIIT IN111421 and IG100223 grants, the Sistema Nacional de Investigadores of CONACyT, and from the CONACyT project number 86372 of the `Ciencia de Frontera 2019’ program, entitled `Citlalc\'oatl: A multiscale study at the new frontier of the formation and early evolution of stars and planetary systems’, M\'exico.

\section*{Data Availability}

No new data were analysed in support of this research.
The data used were already published in \citet{Bus13}.


\bibliographystyle{mnras}
\bibliography{nh3_radial}


\appendix

\section{Radiative transfer calculation}
\label{appendix:model}

In the following we will derive the intensities of the \nh{}
$(1,1;\mm)$, $(1,1;\ms)$, and $(2,2;\mm)$ lines for a spherical envelope by integration of the radiative transfer equation, and calculate the homogeneous optical depths $\tau\sub{hom}(1,1;\ms)$ and 
$\tau\sub{hom}(2,2;\ms)$, and the homogeneous temperature $\Th$ derived from the homogeneous analysis.

\subsection{The radiative transfer equation}

The radiative transfer equation is evaluated for every projected radius $b$. 
Let us consider the slab of geometrical width $\Delta z$, and optical depth $\Delta\tau$, located at position $z$ along the line of sight (see Fig.\ \ref{fig:geom}), with $z^2+b^2=r^2$. 
For this elementary slab with optical depth $\Delta\tau_\nu$, 
the incoming intensity at $z$ is attenuated a factor $\exp(-\Delta\tau_\nu)$, and is increased by the average source function of the slab, times $[1-\exp(-\Delta\tau)]$,
\begin{equation}\label{eq:inu}
I_\nu(z+\Delta z)= I_\nu(z)\exp(-\Delta\tau)+S_\nu(\bar{z}) [1-\exp(-\Delta\tau)],
\end{equation}
where $\bar{z}$ can be approximated by the center of the slab, $\bar{z}=z+\Delta z/2$.
By subtracting $I_\nu(z)$ on both sides of the equation and rearranging terms the equation can be written as
\begin{equation}\label{eq:rte}
\Delta I_\nu= [S_\nu(\bar{z})-I_\nu(z)][1-\exp(-\Delta\tau)].
\end{equation}
The source function $S_\nu$ is defined by the excitation temperature $\Tex$
\begin{equation}
S_\nu=\frac{2h\nu^3}{c^2}\, \frac{1}{\exp(h\nu/k\Tex)-1}=
\frac{2k\nu^2}{c^2} J_\nu(\Tex),
\end{equation}
Equation \ref{eq:rte} can be written in terms of the brightness temperature $\Tb$ (giving the intensity) and the excitation temperature $\Tex$ (giving the source function),
\begin{equation}\label{eq:jnu}
\Delta J_\nu(\Tb)= [J_\nu(\Tex)(\bar{z})-J_\nu(\Tb)(z)]
[1-\exp(-\Delta\tau)].
\end{equation}

\subsection{Excitation temperature $\Tex$}

Assuming that the excitation temperature of the $(1,1)$ transition is given by the two-levels model, $\Tex$ can be expressed as a weighted average of the kinetic and brightness background temperatures, $\Tk$ and $\Tbg$,
\begin{equation}\label{eq:2lev}
J_\nu(\Tex)= \frac{J_\nu(\Tk)+y J_\nu(\Tbg)}{1+y},
\end{equation}
where the weight factor $y$ is
\begin{equation}\label{eq:y}
y=\frac{A_{11}}{n\gamma_{11}}\, \frac{1}{1-e^{-h\nu/k\Tk}},
\end{equation}
$A_{11}$ is the spontaneous emission coefficient,
$n$ is the hydrogen density, $n(\mathrm{H}_2)$, and  
$\gamma_{11}$ is the de-excitation collisional coefficient, 
\begin{equation}
\left[\frac{\gamma_{11}}{\mathrm{s^{-1} cm^3}}\right]=
2.27\times10^{-11}\,
\left[\frac{\Tk}{\mathrm{K}}\right]^{1/2}.
\end{equation}
Thus, for the slab of material between $z$ and $z+dz$ we consider $\Tk=T(r)$, and a background temperature given by the brightness temperature of the incident radiation from the preceding slab, $\Tbg=\Tb(z)$. 
Then, the excitation temperature of the $(1,1)$ transition for this slab is obtained from Eq.\ \ref{eq:2lev}. 
For the $(2,2)$ transition we take the same excitation temperature.

\subsection{Optical depth $\tau$}

Once the excitation temperature is known, we can estimate the elementary optical depths of the $(1,1;\mm)$,  $(1,1;\ms)$, and $(2,2;\mm)$ lines for the slab,
\begin{equation}\label{eq:dtau2}
\Delta\tau(j,k;l) = \sqrt{\frac{4\ln2}{\pi}}
\frac{c^3A_{jk}R_{jkl}}{8\pi\nu_{jk}^3\Delta v} \,
\frac{\exp{(h\nu_{jk}/k\Tex)}-1}{\exp{(h\nu_{jk}/k\Tex)}+1} \,
n_{jk}\, \Delta z, 
\end{equation}
with the volume densities $n_{11}$ and $n_{22}$ given by
\begin{eqnarray}
n_{11} \equal f_{11}(T(r))\, X\, n(r)\\
n_{22} \equal f_{22}(T(r))\, X\, n(r)
\end{eqnarray}
where the fractional abundances $f_{11}$ and $f_{22}$ are given by Eqs.\ \ref{eq:f11f22} and \ref{eq:f22}, with the rotational temperature calculated from the kinetic temperature (Eq.\ \ref{eq:tk_trot}).

\begin{figure}
\centering
\includegraphics[width=0.8\columnwidth]{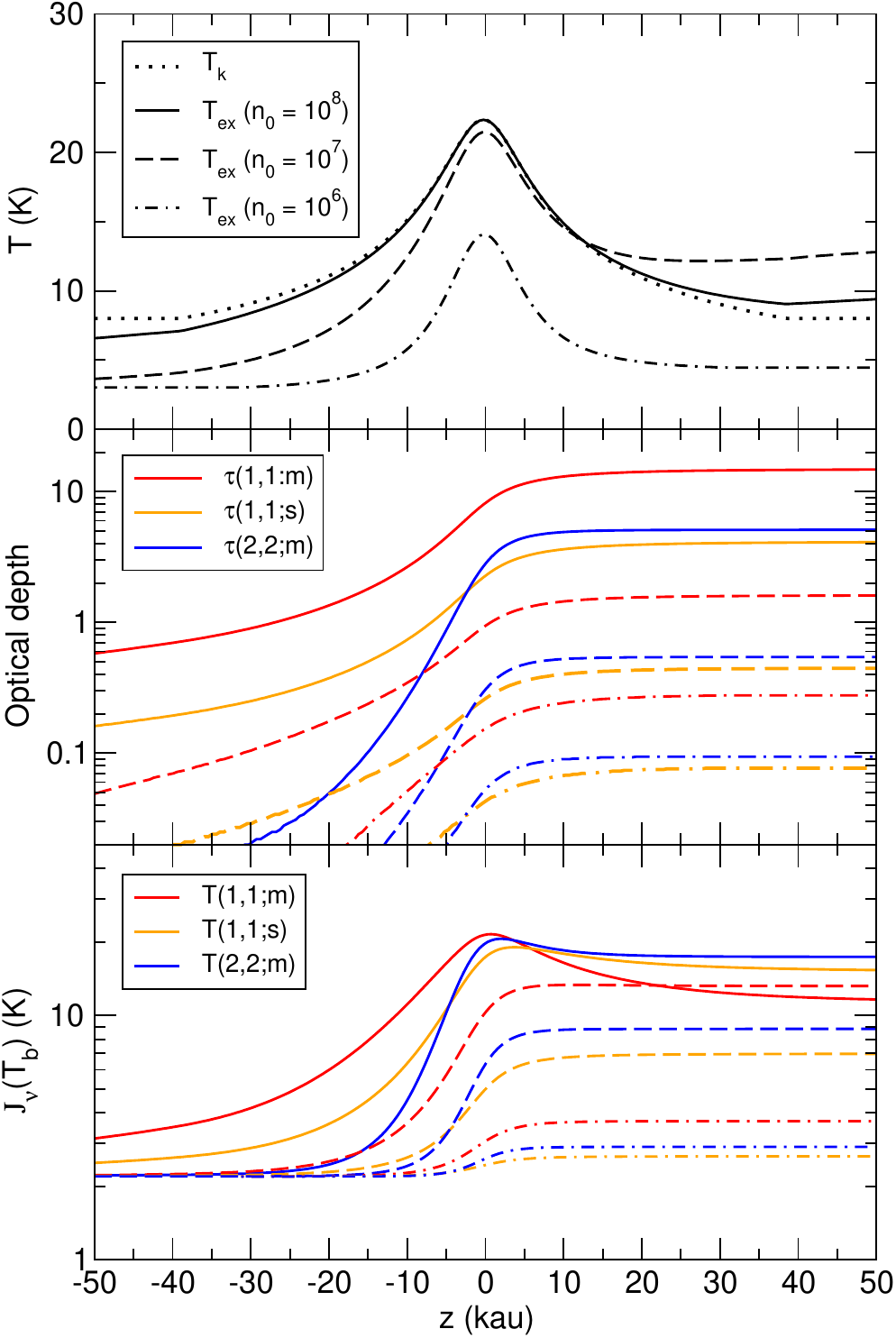}
\caption{\label{fig:thom_los}
Examples of integration along the line of sight. The observer is located rightwards. The distance is 1 kpc, and the projected radius $b$ is $5''$ (5000 au, or 5 kau). 
The envelope parameters are
$p= 2$, $r_c= 0$,
$T_0= 50$ K, $q= 0.5$, 
$\Delta v= 0.5$ \kms{},
$n\sub{min}=10^3$ \cmt{}, 
$T\sub{min}=8$ K, 
\nh{} abundance $X= 10^{-8}$
and three values of the density,
$n_0= 10^8$ (continuum lines), $10^7$ (dashed lines), and $10^6$ \cmt{} (dash-dotted lines).
\textit{Top:} Temperature of the envelope, $\Tk$, and 
excitation temperatures $\Tex$.
\textit{Middle:} Optical depths 
$\tau(1,1;\mm)$ (red), 
$\tau(1,1;\ms)$ (orange), and 
$\tau(2,2;\mm)$ (blue).
\textit{Bottom:} Line intensities 
$J_\nu(\Tb)(1,1;\mm)$ (red),  
$J_\nu(\Tb)(1,1;\ms)$ (orange), and 
$J_\nu(\Tb)(2,2;\mm)$ (blue). 
}
\end{figure}

\subsection{Iterative scheme}

For any given projected radius $b$, the integral along the line of sight has to be evaluated from $-z\sub{max}$ to $z\sub{max}$, where $z\sub{max}=\sqrt{r\sub{env}^2-b^2}$.

The initial conditions at $z=-z\sub{max}$ are 
$\tau(1,1;\mm)= \tau(1,1;\ms)= \tau(2,2;\mm)= 0$,
$\Tb(1,1;\mm)= \Tb(1,1;\ms)= \Tb(2,2;\mm)= \Tbg= 2.752$~K.

For a given $z$, we have determined from the preceding step the values at $z$ of
the optical depths $\tau(1,1;\mm)$, $\tau(1,1;\ms)$, $\tau(2,2;\mm)$, and
the brightness temperatures of the lines, $\Tb(1,1;\mm)$, $\Tb(1,1;\ms)$, $\Tb(2,2;\mm)$.

For this step, from $z$ to $z+\Delta z$, we take
$\bar{z}= z+\Delta z/2$ \citep[this is equivalent to the second order Runge-Kutta method,][]{Abra72}, 
$r=\sqrt{b^2+\bar{z}^2}$,
Eq.\  \ref{eq:2lev} is used with $\Tk= T(r)$, $\Tbg= \Tb(1,1;\mm)$, and $n =n(r)$ to calculate $\Tex$.
The densities $n_{11}$ and $n_{22}$ are calculated, $n_{jk}= f_{jk}\, X\, n(r)$, and Eq.\  \ref{eq:dtau2} is used to estimate
$\Delta\tau(1,1;\mm)$, $\Delta\tau(1,1;\ms)$, and $\Delta\tau(22;\mm)$.
Finally, Eq.\  \ref{eq:jnu} is used to calculate 
$\Delta J_\nu(\Tb(1,1;\mm))$, $\Delta J_\nu(\Tb(1,1;\ms))$, and $\Delta J_\nu(\Tb(2,2;\mm))$.

An example of the integration along the line of sight is shown in Fig.\ \ref{fig:thom_los}, for a power-law density with three different values of the density. For $n_0=10^6$ \cmt{} the three lines, $(1,1;\mm)$, $(1,1;\ms)$, and $(2,2;\mm)$ are optically thin; 
for $n_0=10^7$ \cmt{} the $(1,1;\mm)$ line is moderately optically thick; and for $n_0=10^8$ \cmt{} the three lines are optically thick, the $(1,1;\mm)$ line being very thick (middle panel). 
In the latter case we see that the three lines present self-absorption from the colder gas of the part of the envelope facing the observer, for $z>0$ (bottom panel).
For the (1,1:m) line, optically thicker than the others, the self-absorption is higher, resulting in an intensity lower than those of the $(1,1;\ms)$ and $(2,2;\mm)$ lines.

The final step is to derive the intensities of the lines,
$\Tl(1,1;\mm)= J_\nu(\Tb(1,1;\mm))-J_\nu(\Tbg)$,
$\Tl(1,1;\ms)= J_\nu(\Tb(1,1;\ms))-J_\nu(\Tbg)$, and
$\Tl(2,2;\mm)= J_\nu(\Tb(2,2;\mm))-J_\nu(\Tbg)$.
Once the three line intensities are known, the standard analysis of Sec.\  \ref{sec:basics} is used to derive the homogeneous estimates of the optical depths $\tau\sub{hom}(1,1;\mm)$, $\tau\sub{hom}(1,1;\ms)$, and $\tau\sub{hom}(2,2;\mm)$, the homogeneous estimate of the rotational temperature, and finally the homogeneous temperature, $\Th$.

\begin{figure}
\centering
\includegraphics[width=0.85\columnwidth]{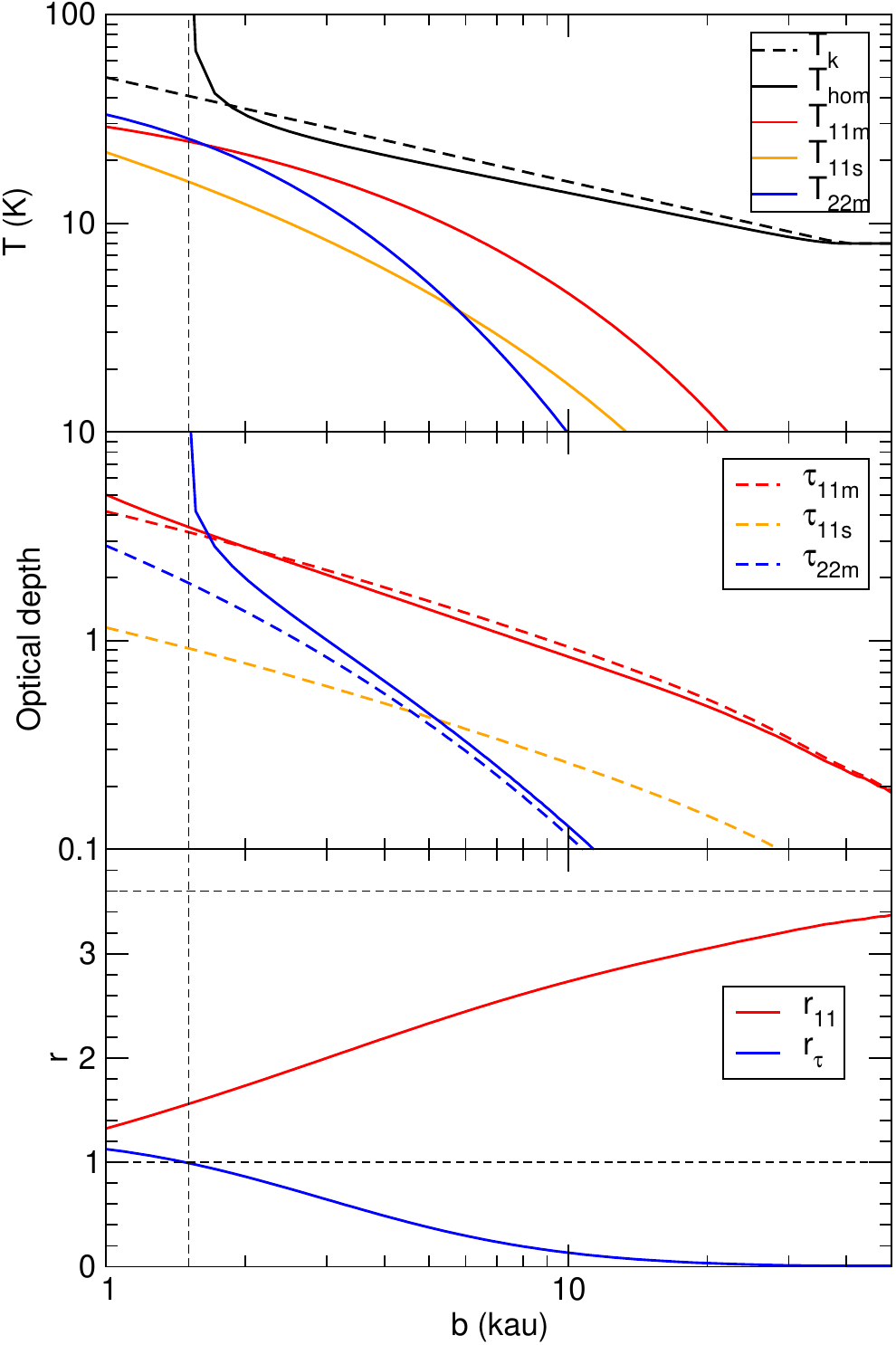}
\caption{
Example of radial profiles, obtained for an envelope at a distance of 1 kpc, with
$n_0= 10^7$ \cmt{}, $p= 2$, $r_c= 0$,
$T_0= 50$ K, $q= 0.5$,
$\Delta v= 0.5$ \kms{},
$n\sub{min}= 10^3$ \cmt{}, 
$T\sub{min}= 8$ K, and 
\nh{} abundance $X= 10^{-8}$.
\textit{Top panel:} 
Temperature of the envelope (black dashed line),
line intensities
$\Tl(1,1;\mm)$ (red),  
$\Tl(1,1;\ms)$ (orange),  
$\Tl(2,2;\mm)$ (blue), 
and homogeneous temperature $\Th$ (black solid line), for the values of $b$ for which it could be calculated.
\textit{Central panel:} Optical depths 
$\tau(1,1;\mm)$ (red), 
$\tau(1,1;\ms)$ (orange), and 
$\tau(2,2;\mm)$ (blue).
The dashed lines are the values obtained from the integration along the line of sight. The continuum lines are the values obtained from the homogeneous analysis.
\textit{Bottom panel:}
$r_{11}= T_{1\mm}/T_{1\ms}$ (red), and
$r_\tau= [1-\exp(-\tau_{1\mm})]/[1-\exp(-a\,\tau_{1\mm})] r_{12}$ (blue),
where 
$r_{12}=T_{2\mm}/T_{1\mm}$ and
$a= 3.544$ (see text).
The horizontal dashed lines at 1 and 3.6 indicate the minimum and maximum values of $r_{11}$ allowed for the homogeneous analysis (see Section \ref{sec:limitations}). 
The vertical dashed line at $r_\tau=1$ indicates the minimum projected radius $b$ for which the homogeneous temperature can be obtained.
\label{fig:thom}
}
\end{figure}

\subsection{Results}

In Fig.\ \ref{fig:thom} we show an example of results obtained for an envelope with a power-law density profile, at a distance of 1 kpc. 
For this distance, an angular size of $1''$ corresponds to a linear size of $1 \mbox{ kau} = 1000$ au. 
The parameters of the envelope are a
reference radius $r_0= 1$ kau,
density parameter $n_0=10^7$ \cmt{}, 
density power-law index $p= 2$, 
radius of the inner flat region $r_c=0$,
temperature at $r_0$, $T_0= 50$ K, 
temperature power-law index $q= 0.5$, 
line width $\Delta v= 0.5$ \kms{},
minimum density (setting the envelope radius) $n\sub{min}=10^3$ \cmt{}, 
minimum temperature $T\sub{min}=8$ K,
and \nh{} abundance $X= 10^{-8}$.
The top panel shows the radial profiles of 
temperature of the envelope (dashed line), 
line intensities, 
$\Tl(1,1;\mm)$, 
$\Tl(1,1;\ms)$,  
$\Tl(2,2;\mm)$, and 
homogeneous temperature $\Th$.
The middle panel shows the radial profiles of optical depths of the lines,
$\tau(1,1;\mm)$, 
$\tau(1,1;\ms)$, and 
$\tau(2,2;\mm)$, 
calculated from the integration along the line of sight (dashed lines), and estimated from the homogeneous analysis (continuum lines).

As can be seen in the top panel, for small values of the projected radius the homogeneous temperature $\Th$ could not be derived. The reason is that for $b$ below a critical value of 1.94 kau, the intensity of the $(2,2)$ line is too high to fulfil the condition of Eq.\  \ref{eq:cond_rtau}, and $\tau\sub{hom}(2,2;\mm)$ could not be calculated (see bottom panel of Fig.\ \ref{fig:thom}). 
The failure of the homogeneous analysis is a consequence of the $(1,1;\mm)$ line being optically thick for small $b$, and thus tracing the cold part of the envelope facing the observer, while the $(2,2;\mm)$ line, with a lower optical depth, is tracing hotter gas near the center of the envelope.
For values of the projected radius near the critical value given above, the optical depth of the $(2,2)$ line derived from the homogeneous analysis is much higher than the actual optical depth, and the homogeneous temperature can be very different from the envelope temperature.
For larger projected radii, far from the critical value, the radial profile of the homogeneous temperature is similar to, but systematically below that of the envelope temperature.

\subsection{Comparison with LIME}
\label{sec:lime}

\begin{figure}
\centering
\includegraphics[width=0.85\columnwidth]{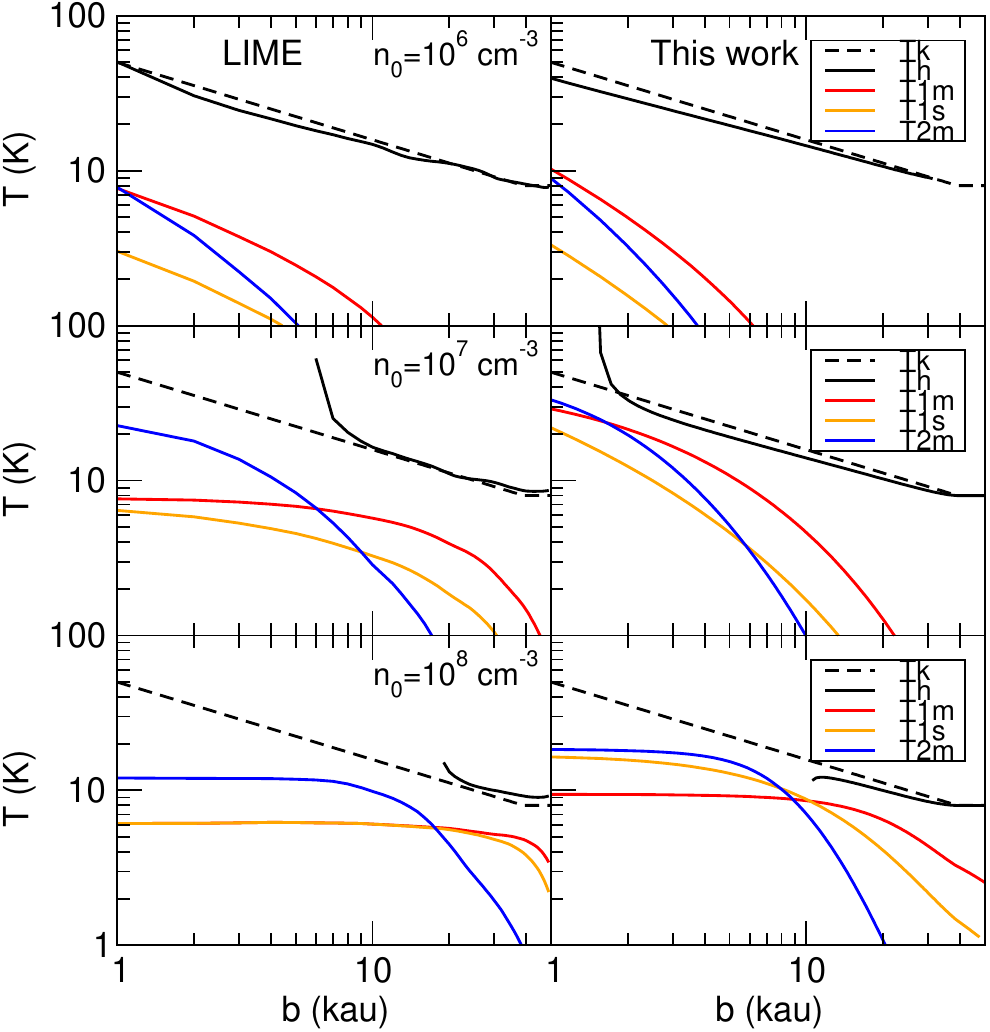}
\caption{
Comparison of LIME (left) and radiative transfer of this work (right) of the radial profiles of the 
temperature of the envelope (black dashed line),
line intensities
$\Tl(1,1;\mm)$ (red),  
$\Tl(1,1;\ms)$ (orange),  
$\Tl(2,2;\mm)$ (blue), 
and homogeneous temperature $\Th$ (black solid line).
The envelope parameters are a
reference radius $r_0= 1$ kau,
density parameter $n_0=10^6$ \cmt{} (top), $10^7$ \cmt{} (middle), and $10^8$ \cmt{} (bottom), 
density power-law index $p= 2$, 
radius of the inner flat region $r_c=0$,
temperature at $r_0$, $T_0= 50$ K, 
temperature power-law index $q= 0.5$, 
line width $\Delta v= 0.5$ \kms{},
minimum density $n\sub{min}=10^3$ \cmt{}, 
minimum temperature $T\sub{min}=8$ K,
and \nh{} abundance $X= 10^{-8}$.
\label{fig:lime}}
\end{figure}

The Line Modelling Engine (LIME) \citep{Bri10} was used to compare the spherical envelope model with a full radiative transfer calculation.
LIME assumes that the gas and dust temperatures are the same.
Unfortunately, the collision coefficients of the quadrupole-hyperfine transitions of \nh{} are not available in the literature. 

LIME was run using the molecular data for para-\nh{} from the Leiden Atomic and Molecular Database (LAMDA) \citep{Sch05}, which contains energy levels, transition frequencies, Einstein coefficients of the inversion transitions, and the collision rates with para-H$_2$.
The output of LIME provided us the radial profiles of the intensities and optical depths of the inversion transitions \nh{} $(1,1)$ and $(2,2)$, $\Tl(j,k)$ and $\tau(j,k)$.
The intensity of the main and inner satellite quadrupole-hyperfine lines was then estimated as
\begin{eqnarray}\label{eq:lime}
\Tl(j,k;\ms) \equal \Tl(j,k)\, \frac{1-\exp[-R_{jks}\,\tau(j,k)]}{1-\exp[-\tau(j,k)]}, \\
\Tl(j,k;\mm) \equal \Tl(j,k)\, \frac{1-\exp[-R_{jkm}\,\tau(j,k)]}{1-\exp[-\tau(j,k)]}. \nonumber
\end{eqnarray}
From these intensities, the standard \nh{} analysis was performed to obtain the homogeneous temperature radial profile.

The LIME calculation assumes that the five quadrupole-hyperfine lines of each inversion transition are completely blended, with an optical depth that is the total optical depth of the inversion transitions.
So, the inversion optical depths calculated by LIME are higher than the actual optical depths of the hyperfine lines. 
As a consequence, for partially thick emission, the region contributing to the LIME intensity is closer to the observer (that is, farther from the envelope center, and so cooler and less dense) than for the hyperfine intensity.
Thus, we expect discrepancies between LIME calculations and the radiative transport calculations of this work, which could be more important in the optically thick regime.

In the left panels of Fig.\ \ref{fig:lime} we show the line intensities $\Tl(1,1;\mm)$, $\Tl(1,1;\ms)$, $\Tl(2,2;\mm)$, and homogeneous temperature $\Th$ for three different density parameters, $n_0= 10^6$, $10^7$, and $10^8$ \cmt{} calculated with LIME.
In all three cases the homogeneous temperatures obtained from the the LIME calculation, at the projected radii where it can be calculated (solid black line), are near the envelope temperature. 

In the right panels of Fig.\ \ref{fig:lime} we show, for comparison, the same results calculated with the spherical envelope model of this work. 
As can be seen, at low densities ($n_0= 10^6$ \cmt{}) there is a good agreement between the line intensities calculated by both methods, while at higher densities the discrepancies are most probably due to the overestimation of the optical depths made by LIME, as discussed above.
Thus, despite the shortcomings of the LIME calculations, there is a general agreement in the line intensities and, more important, a good agreement between the homogeneous temperature \Th{} derived from both methods.
In conclusion, we can say that the radiative transfer calculation used in the present work is correct.


\bsp	
\label{lastpage}
\end{document}